\pdfoutput=1  % Instructs arXiv to run this with pdflatex rather than latex.

\documentclass[aps,twocolumn,groupedaddress,floatfix,preprintnumbers,nofootinbib,eqsecnum]{revtex4}
\usepackage{amssymb,amsmath,placeins,graphicx,multirow}

\begin{document}

\title{Dynamical Dark Matter from Thermal Freeze-Out}
\author{Keith R.\ Dienes$^{1,2}$\footnote{E-mail address:  {\tt dienes@email.arizona.edu}},
      Jacob Fennick$^{3}$\footnote{E-mail address:  {\tt jfennick@hawaii.edu}},
      Jason Kumar$^{4}$\footnote{E-mail address:  {\tt jkumar@hawaii.edu}}, 
      Brooks Thomas$^{5}$\footnote{E-mail address:  {\tt thomasbd@lafayette.edu}} 
      }
\affiliation{
     $^1\,$Department of Physics, University of Arizona, Tucson, AZ  85721  USA\\
     $^2\,$Department of Physics, University of Maryland, College Park, MD  20742  USA\\
     $^3\,$Department of Mathematics, University of Hawaii, Honolulu, HI 96822  USA\\
     $^4\,$Department of Physics \& Astronomy, University of Hawaii, Honolulu, HI 96822  USA\\
     $^5\,$Department of Physics, Lafayette College, Easton, PA  18042  USA}
%  \date{\today}

\begin{abstract}
In the Dynamical Dark Matter (DDM) framework, the dark sector comprises
a large number of constituent dark particles
whose individual masses, lifetimes, and cosmological abundances
obey specific scaling relations with respect to each other.
In particular,
the most natural versions of this framework tend 
to require a spectrum of cosmological abundances which scale
inversely with mass, so that dark-sector states with larger masses
have smaller abundances.
Thus far, DDM model-building has 
primarily relied on non-thermal mechanisms for abundance generation such as 
misalignment production, since these mechanisms give rise to abundances that
have this property.
By contrast, the simplest versions of thermal freeze-out tend to produce
abundances that increase, rather than decrease, with the mass of the dark-matter component.
In this paper, we demonstrate that there exist relatively simple
modifications of the traditional thermal freeze-out mechanism which
``flip'' the resulting abundance spectrum, producing abundances
that scale inversely with mass.
Moreover, we demonstrate that 
a far broader variety of scaling relations between lifetimes, abundances,
and masses can emerge through thermal freeze-out than through the non-thermal
mechanisms previously considered for DDM ensembles.  
The results of this paper thus extend the DDM framework
into the thermal domain and essentially allow us to ``design'' our resulting
DDM ensembles at will in order to realize a rich array of resulting 
dark-matter phenomenologies. 
\end{abstract}

\maketitle

%========================================================================
%          KEYSROKE-SAVING MACROS, nothing complicated 
%========================================================================
\newcommand{\newc}{\newcommand}
\newc{\gsim}{\lower.7ex\hbox{$\;\stackrel{\textstyle>}{\sim}\;$}}
\newc{\lsim}{\lower.7ex\hbox{$\;\stackrel{\textstyle<}{\sim}\;$}}
\makeatletter
\newcommand{\biggg}{\bBigg@{3}}
\newcommand{\Biggg}{\bBigg@{4}}
\newcommand{\tev}{\text{TeV}}
\newcommand{\kev}{\text{keV}}

\makeatother

\def\vac#1{{\bf \{{#1}\}}}

\def\beq{\begin{equation}}
\def\eeq{\end{equation}}
\def\beqn{\begin{eqnarray}}
\def\eeqn{\end{eqnarray}}
\def\calM{{\cal M}}
\def\calV{{\cal V}}
\def\calF{{\cal F}}
\def\half{{\textstyle{1\over 2}}}
\def\quarter{{\textstyle{1\over 4}}}
\def\ie{{\it i.e.}\/}
\def\eg{{\it e.g.}\/}
\def\etc{{\it etc}.\/}
%\def\alpha\dot{\alpha}{\alpha\dot{\alpha}}

%     The following macros are to create the "blackboard bold"
%     characters for "R" (set of real numbers),
%     "C" (set of complex numbers), and "Q" (set of rational numbers).

\def\chibar{{\overline{\chi}}}
\def\psibar{{\overline{\psi}}}
\def\inbar{\,\vrule height1.5ex width.4pt depth0pt}
\def\IR{\relax{\rm I\kern-.18em R}}
 \font\cmss=cmss10 \font\cmsss=cmss10 at 7pt
\def\IQ{\relax{\rm I\kern-.18em Q}}
\def\IZ{\relax\ifmmode\mathchoice
 {\hbox{\cmss Z\kern-.4em Z}}{\hbox{\cmss Z\kern-.4em Z}}
 {\lower.9pt\hbox{\cmsss Z\kern-.4em Z}}
 {\lower1.2pt\hbox{\cmsss Z\kern-.4em Z}}\else{\cmss Z\kern-.4em Z}\fi}
\def\TBBN{T_{\mathrm{BBN}}}
\def\OmegaCDM{\Omega_{\mathrm{CDM}}}
\def\OmegaDM{\Omega_{\mathrm{CDM}}}
\def\Omegatot{\Omega_{\mathrm{tot}}}
\def\rhocrit{\rho_{\mathrm{crit}}}
\def\tnow{t_{\mathrm{now}}}
\def\arcsinh{\mbox{arcsinh}}
\def\Omegatotnow{\Omega_{\mathrm{tot}}^\ast}
\def\mij{m_{jj}}
\def\mijmin{m_{jj}^{(\mathrm{min})}}
\def\mijmax{m_{jj}^{(\mathrm{max})}}
\def\mmax{m_{\mathrm{max}}}
\def\epsig{\epsilon_{\mathrm{sig}}}
\def\Lint{\mathcal{L}_{\mathrm{int}}}
\def\MT2{M_{T2}}
\def\MTtwomax{M_{T2}^{\mathrm{max}}}
\def\BR{\mathrm{BR}}
%This is the old version: the slash is further to the left in the "E" in this version.
%\newcommand{\Dsle}[1]{\hskip 0.09 cm \slash\hskip -0.20 cm #1}
\newcommand{\Dsle}[1]{\hskip 0.09 cm \slash\hskip -0.23 cm #1}
\newcommand{\Dirsl}[1]{\hskip 0.09 cm \slash\hskip -0.20 cm #1}
\newcommand{\met}{{\Dsle E_T}}

\newcommand{\fb}{{\rm fb}}
\newcommand{\ifb}{{\rm fb}^{-1}}
\newcommand{\pb}{{\rm pb}}
\newcommand{\ipb}{{\rm pb}^{-1}}

%========================================================================

\input epsf

%========================================================================
%========================================================================
%               MAIN TEXT BEGINS HERE
%========================================================================

%========================================================================

%\tableofcontents

%%%%%%%%%%%%%%%%%%%%%%%%%%%%%%%%%%%%%%%%%%%%%%%%%%%%%%%%%%%%%%%%%%%%%%%%%%%%%%%%%%%%%%

\section{Introduction\label{sec:Introduction}}

%%%%%%%%%%%%%%%%%%%%%%%%%%%%%%%%%%%%%%%%%%%%%%%%%%%%%%%%%%%%%%%%%%%%%%%%%%%%%%%%%%%%%%

Dynamical Dark Matter (DDM)~\cite{DDM1,DDM2} is a framework for dark-matter
physics in which the dark sector is composed of a large ensemble of dark states
exhibiting a variety of masses, lifetimes, and cosmological abundances.
The phenomenological viability of this framework rests 
on a balancing between the lifetimes and abundances of the
individual ensemble constituents, so that
states with larger abundances have
longer lifetimes while states with smaller abundances can have correspondingly smaller lifetimes.
Such a balancing is required in order to satisfy observational 
constraints on dark-matter decay.  

Scenarios within the DDM framework give rise to distinctive signatures at 
colliders~\cite{DDMLHC1,DDMLHC2}, at direct-detection
experiments~\cite{DDMDD}, and at indirect-detection experiments~\cite{DDMAMS,cosmicray1,cosmicray2}.  
Such scenarios also give rise to enhanced complementarities~\cite{DDMComp,DDMComp2} between different
types of experimental probes.
Moreover, DDM ensembles have been shown to arise naturally in a number of 
scenarios for new physics beyond the Standard Model (SM).  These include theories with extra spacetime 
dimensions~\cite{DDM1,DDM2,DDMAxion}, theories involving strongly-coupled
hidden sectors~\cite{DDMHadrons}, theories involving large spontaneously-broken 
symmetry groups~\cite{RandomMatrixDDM},
and even string theories~\cite{DDMHadrons,anupam}.  
In these and other realistic DDM scenarios, the masses, lifetimes, 
and abundances of these individual particles are not arbitrary.  
Rather, these quantities follow directly from the underlying physics model 
 and generally take the form of {\it scaling relations}\/
which dictate
how these quantities scale relative to one another across the ensemble as a 
whole.  Through these scaling relations, the properties of the 
ensemble constituents --- and thus the properties of the ensemble itself --- 
are completely specified through only a small number of free parameters.     
Thus, even though the number of particles which 
contribute to the total dark-matter abundance is typically quite large,
DDM scenarios of this sort are every bit as predictive as traditional 
dark-matter scenarios. 

One of the most fundamental of these scaling relations 
is that describing the relationship between the masses of the individual ensemble 
constituents and their cosmological abundances.  This scaling relation in turn depends 
crucially on the mechanism through which the abundances for these constituents are
generated.  Thus, the properties of this scaling relation depend not only on the underlying 
 {\it particle-physics model}\/, but also on the underlying {\it cosmological history}\/ in
which it is embedded.
For example, in DDM models in which the ensemble constituents are the Kaluza-Klein (KK) 
modes of an axion-like particle propagating in the bulk of a theory with extra 
spacetime dimensions, misalignment production provides a natural 
abundance-generation mechanism for these constituents~\cite{DDM1,DDM2,DDMAxion}.  
Likewise, in DDM models in which the ensemble constituents are composite
states in the confining phase of a strongly-coupled hidden sector, it turns out
that residual 
gauge interactions at temperatures just below the confinement scale give rise to an
appropriate spectrum of abundances which compensates for the exponential rise in
the density of states~\cite{DDMHadrons}.  
However, these abundance-generation mechanisms are only compatible with particular 
classes of particle-physics models.  It is therefore useful to explore alternative 
mechanisms for abundance generation --- mechanisms which might be applicable in a 
broader variety of DDM contexts.   

Of course, one of the most widely discussed and widely exploited methods of abundance
generation in the dark-matter literature is
thermal freeze-out
(for reviews, see, \eg, 
Refs.~\cite{KolbAndTurner,GondoloGelmini,JungmanKamionkowskiGriest,Dodelson}).
Indeed, thermal freeze-out provides
a natural and versatile mechanism through which a neutral, weakly-interacting massive particle (WIMP) species
$\chi$ which is initially in thermal equilibrium can acquire a present-day abundance 
$\Omega_\chi$ on the order of the total present-day dark-matter abundance 
$\OmegaDM \approx 0.26$~\cite{Planck}.  Indeed, this mechanism not only underpins 
the so-called ``WIMP miracle,'' but also generically yields $\Omega_\chi\sim \OmegaDM$ 
for a broader class of dark-matter particles which do not participate in 
SM weak interactions but which nevertheless have annihilation 
cross-sections similar to that of a traditional WIMP~\cite{WIMPless}.  
The range of dark-matter masses $m_\chi$ for which the freeze-out is typically relevant is 
$\mathcal{O}(1\mbox{~keV}) \lesssim m_\chi \lesssim \mathcal{O}(100\mbox{~TeV})$. 
The lower limit to this range stems from the requirement that the dark-matter 
candidate be ``cold'' --- \ie, non-relativistic --- during the freeze-out epoch
(see, \eg, Ref.~\cite{LowerBoundFreezeoutMass}),
while the upper limit stems from considerations related to perturbative 
unitarity~\cite{GriestUnitarity}.  However, there are ways of circumventing 
this upper bound and broadening the window of applicability.  For example, this 
bound is considerably relaxed in theories in which the dark and hidden sectors 
thermally decouple well before the freeze-out epoch~\cite{AsherDefiesGriest}.   

The question then arises as to whether thermal freeze-out can yield a spectrum
of cosmological abundances that are suitable for a DDM ensemble. 
At first glance, it may seem that this is not possible.
The reason is relatively simple.
In a DDM ensemble, the abundances of the ensemble constituents must generally scale 
inversely to their decay widths.
However, the decay widths of such states generally scale as a positive power
of the mass.   
This then requires the
cosmological abundances of the ensemble constituents to scale {\it inversely}\/ with their masses:
\beq
            \Omega_i ~\sim~ m_i^\gamma ~~~~{\rm where} ~\gamma<0~.
\eeq
Unfortunately, 
while this holds for all of the non-thermal production mechanisms that have thus far
been exploited for DDM abundance generation,
this is generally not a property of thermal freeze-out.
Indeed, as we know, the WIMP miracle itself rests upon the classic observation 
that~\cite{ZeldovichFreezeout,ChiuFreezeout,SteigmanFreezeout,ScherrerTurner,WIMPless}
\beq
            \Omega_\chi  ~\sim~  {m_\chi^2\over g_\chi^4}~,
\label{WIMPmiracle}
\eeq
implying the canonical value $\gamma = +2$.
Thus, all else being equal, dark-matter particles with larger masses can be 
expected to retain larger cosmological abundances 
after freeze-out than those with smaller masses ---
precisely the {\it opposite}\/ of what is generically needed for a DDM ensemble.
 
In this paper, we shall demonstrate that an acceptable spectrum of abundances can nevertheless be 
generated for a DDM ensemble through thermal freeze-out, with abundances $\Omega_i$ scaling
 {\it inversely}\/ with masses $m_i$ across the ensemble.
Indeed, this can occur even if the couplings $g_i$ are universal
across all ensemble constituents.
Moreover, as we shall demonstrate, such thermal freeze-out scenarios can 
arise completely naturally, without any fine-tuning.
Indeed, we shall find that such scenarios can even give rise
to a wide variety of possible scaling behaviors with a wide range
of possible (negative) scaling exponents $\gamma$. 
Thus, from a model-building perspective,
we shall find that thermal freeze-out actually provides a versatile tool for 
``designing'' viable DDM ensembles with different scaling behaviors and exploring their resulting
phenomenologies.

This paper is organized as follows.  In Sect.~\ref{sec:ThermalFreezeout},  we examine the
ways in which the cosmological abundance of a particle produced by thermal
freeze-out depends on the mass of that particle.  We review how the canonical
relationship between abundance and mass arises within the WIMP
paradigm and illustrate how this
relationship can be altered through modifications of the particle physics
alone, without any modification of the background cosmology.
In Sect.~\ref{general},  we then undertake a somewhat more general study along
the lines of this approach and
derive a general expression for the freeze-out cosmological abundance of
an individual ensemble constituent 
as a function of the mass, spin, and couplings of the particles involved. 
In this way we find that we can generate a broad range of negative scaling exponents $\gamma$
and potentially even dial between them. 
In Sect.~\ref{sec:DecayRateScaling}, we then examine how and under what conditions 
a suitable balancing of decay widths against abundances --- a balancing which is the hallmark
of the DDM framework --- can naturally arise in thermal DDM scenarios.
Finally, in Sect.~\ref{sec:Conclusions},  we conclude with a summary of the
implications of our results for DDM model-building in a thermal context and possible
directions for future work.

%%%%%%%%%%%%%%%%%%%%%%%%%%%%%%%%%%%%%%%%%%%%%%%%%%%%%%%%%%%%%%%%%%%%%%%%%%%%%%%%%%%%%%

\section{Flipping the Abundance Spectrum:~  Integrating Out Prior to Freezing Out
\label{sec:ThermalFreezeout}}

%%%%%%%%%%%%%%%%%%%%%%%%%%%%%%%%%%%%%%%%%%%%%%%%%%%%%%%%%%%%%%%%%%%%%%%%%%%%%%%%%%%%%%

In general, thermal freeze-out of a given dark-matter particle $\chi$ 
results from a competition between $\chi$ annihilation and
the Hubble expansion of the universe.
{\it A priori}\/, we can imagine an annihilation process of the form
$\chibar\chi\to \psibar\psi$ where $\psi$ denotes SM states or even states in the dark sector which are
less massive than $\chi$. 
As the universe expands, the efficient annihilation and production of $\chi$ via this process 
and its reverse ensures
that $\chi$ remains in thermal equilibrium
with $\psi$ and all other particles that are in thermal equilibrium with $\psi$. 
Once the universe has cooled to a point at which $\chi$ is non-relativistic, 
this thermal equilibrium causes $n_\chi$, the particle density  of $\chi$, to fall exponentially as a function of time,
which in turn causes the annihilation rate to fall as well.
This situation persists until 
the annihilation rate $\Gamma$ falls below the Hubble parameter $H$.
At this point, the expansion of the universe has caused $n_\chi$ 
to fall so low that the dark-matter particles 
can no longer efficiently find each other in order to annihilate.
The efficient annihilation and production of $\chi$ then ends,
with $\chi$ falling out of chemical equilibrium and the number of $\chi$ particles remaining essentially
constant thereafter.
In other words, $\chi$ has experienced thermal freeze-out.

\begin{figure}[t]
\centering
\includegraphics[width=0.49\textwidth,keepaspectratio]{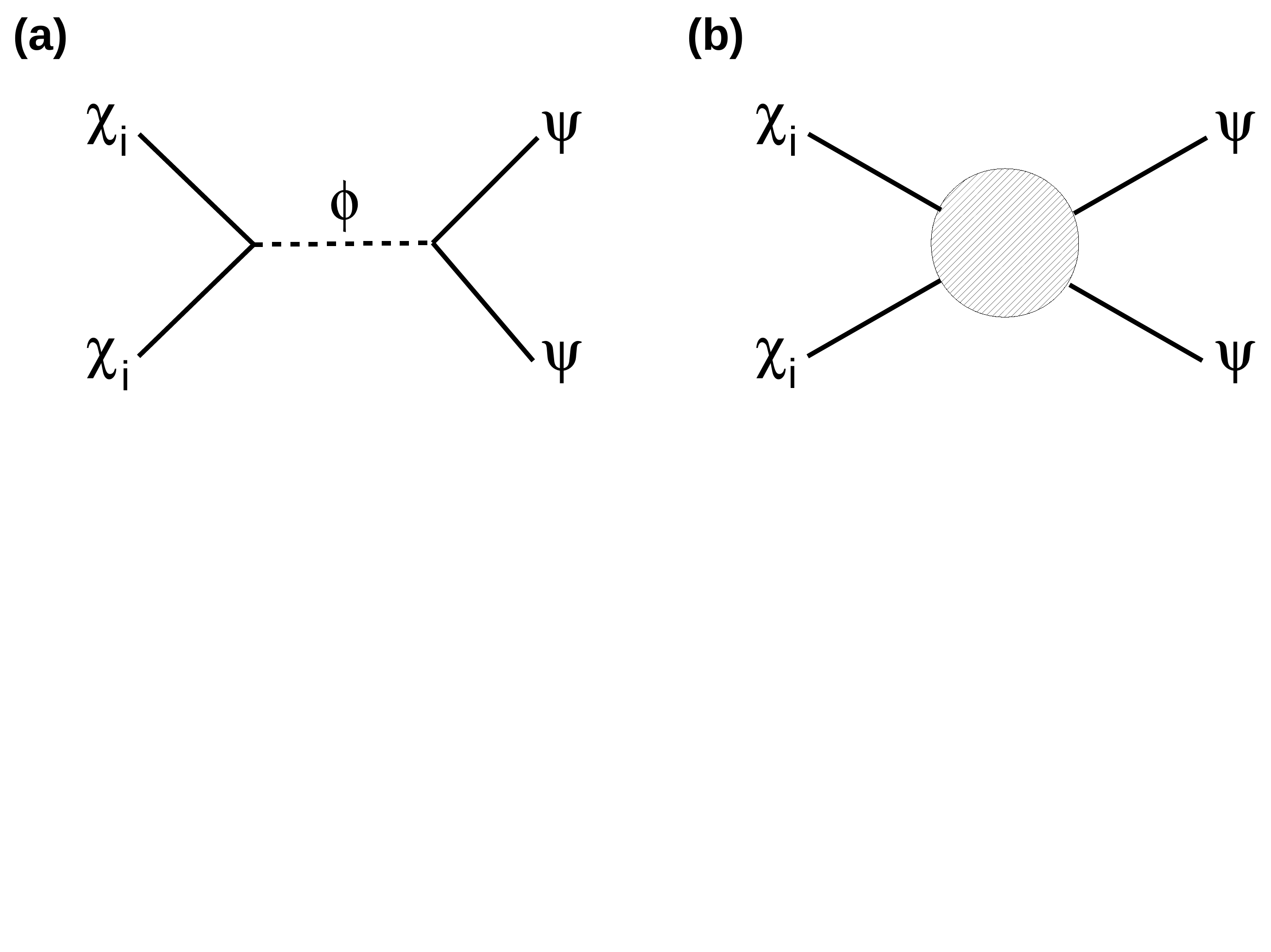}
\vskip -1.2 truein
\caption{Diagrams for dark-matter annihilation $\chibar\chi\to \psibar\psi$.  (a)  
An $s$-channel diagram in which 
annihilation proceeds through a mediator $\phi$.   (b)  A four-point effective contact
interaction which emerges from diagram (a) upon integrating out the mediator $\phi$.
This diagram represents the limit of diagram (a) in which $m_\phi\gg m_\chi, m_\psi$.
}
\label{fig:diagrams}
\end{figure}

Estimating the resulting post-freeze-out dark-matter abundance $\Omega_\chi$
therefore requires knowledge of the thermally-averaged annihilation cross-section $\langle \sigma v\rangle$.
Calculating this quantity in turn requires a set of specific assumptions
concerning how $\chi$ annihilates into SM states or other relatively light states in the dark sector.
In general, there are many processes which can contribute to the overall annihilation cross-section.
However, for the purposes of this paper, we will concentrate on the relatively simple case in which
this annihilation proceeds through an $s$-channel
diagram such as that shown in Fig.~\ref{fig:diagrams}(a) 
in which two dark-matter particles $\chi$
of mass $m_\chi$ annihilate into two light particles $\psi$ through a mediator $\phi$: 
{\it i.e.}\/, $\overline{\chi}\chi\to\phi\to \overline{\psi}\psi$.
Note that we are not assuming that $\chi$ or $\psi$ are their own antiparticles, nor
are we even specifying the spins of these states.
However, for simplicity, we shall begin by assuming that $m_\chi \gg m_\phi, m_\psi$  so that both $\phi$ and $\psi$ can
be taken as effectively massless.  We shall likewise take $g_\chi$ and $g_\psi$ to be constants representing the 
couplings of $\chi$ to $\phi$ and $\phi$ to $\psi$, respectively.
We shall also take each of our incoming dark-matter particles to be non-relativistic, 
with an energy $E_\chi\approx m_\chi$.
It then immediately follows via dimensional analysis that 
our thermally averaged cross-section generically takes the 
form~\cite{ZeldovichFreezeout,ChiuFreezeout,SteigmanFreezeout,ScherrerTurner,WIMPless}
\beq
          \langle \sigma v \rangle ~\sim~  \frac{g_\chi^2 g_\psi^2}{ m_\chi^2}~
\label{WIMPscaling}
\eeq
where $v$ denotes the relative velocity of the dark-matter particles.
As we shall discuss below, 
under rather broad assumptions
the process of thermal freeze-out leads to a residual abundance $\Omega_\chi$ which scales
as $\langle \sigma v\rangle^{-1}$.
This then reproduces the traditional ``WIMP miracle'' result 
\beq
         \Omega_\chi ~\sim~ \frac{m_\chi^2}{g_\chi^2 g_\psi^2}~,
\label{tradresult}
\eeq
thereby yielding the expected scaling behavior $\Omega_\chi\sim m_\chi^\gamma$ with $\gamma= +2$.
This behavior can be realized in a number of other ways as well.

How, then, can we ``flip'' this result and realize a scaling behavior in which $\Omega_\chi\sim m_\chi^\gamma$ with $\gamma<0$?
Note that 
prior to utilizing the non-relativistic
approximation $E\approx m_\chi$ in the process of deriving Eq.~(\ref{WIMPscaling}),
the powers of $m_\chi$ that appear in
Eq.~(\ref{WIMPscaling})
had originally been
powers of $E$.
Thus our interest is really in changing the powers of energy associated with the 
annihilation process.
In particular, we are interested in finding a way to {\it decrease}\/ the powers of energy in our expression for
the abundance $\Omega_\chi$, or equivalently to {\it increase}\/ the powers of energy in the cross-section 
$\langle \sigma v\rangle$.

Of course, one way of changing the powers of energy is already well known:  if we imagine that the mediator $\phi$
has a non-zero mass $m_\phi$, then there is a natural process we may follow which amounts to replacing
\beq
        \frac{g_\chi^2 g_\psi^2}{E^2} ~~~~\longrightarrow ~~~~
        G^2 \, E^2
\label{integratingout}
\eeq
where we have introduced the {\it dimensionful}\/ 
effective coupling $G\equiv g_\chi g_\psi /m_\phi^2$. 
Indeed, this is nothing but the process of {\it integrating out}\/ the mediator $\phi$ --- \ie, of taking
$m_\phi \gg E\approx m_\chi$.
This then leaves us with the effective annihilation process illustrated in Fig.~\ref{fig:diagrams}(b). 
Note that the limit $m_\phi \gg m_\chi$ which underpins the integrating out of $\phi$ is
 {\it opposite}\/ to the limit that yields
the traditional result in Eq.~(\ref{WIMPscaling}).
However, there is no conflict in doing this since we are no longer considering $m_\phi$ as parametrically
tied to the
weak scale.

%================ FIGURE ================================================
%================ FIGURE ================================================

As evident in Eq.~(\ref{integratingout}), this  process of integrating out $\phi$ results
in a growing, unbounded cross-section whose unitarization was previously supplied through $\phi$.
As a result, even for finite $m_\phi$, those dark-matter particles $\chi$ 
whose masses are significantly below $m_\phi$
experience 
an effective annihilation cross-section which grows rather than shrinks as a function
of energy.
For such dark-matter particles, standard thermal freeze-out then yields cosmological abundances which decrease,
rather than increase, as a function of $m_\chi$.

To make these observations more explicit,
let us imagine that we have an ensemble of dark-matter components $\chi_i$, $i=1,...,N$,
with masses $m_i$ satisfying $m_{i+1}\geq m_i$ for all $i\leq N-1$,
whose annihilations are dominated by the
process shown in Fig.~\ref{fig:diagrams}(a)
in which $\chi_i$ and its antiparticle $\chibar_i$ (which may or may not be identified with
$\chi_i$ itself) annihilate into a pair of other, lighter particles $\psi$ and $\psibar$ through the
exchange of a common mediator $\phi$.
For concreteness we shall assume 
that $\chi_i$ and $\psi$ are Dirac fermions
and that $\phi$ is a scalar.
Here $\psi$ collectively denotes a SM state or a state in the dark sector which is lighter than the $\chi_i$,
but the identities of $\psi$ and $\psibar$ are not particularly important
for determining the abundance spectrum and we shall not specify their identities further.
Likewise, we shall assume that the couplings between these fields take the simple forms
$g_\chi \chibar_i \chi_i \phi$
and 
$g_\psi \phi \psibar \psi$
where $g_\chi$ and $g_\psi$ are arbitrary coupling constants. 
Note, in particular,  that we are taking $g_\chi$ to be independent of $i$ and hence
universal for the entire ensemble;  other options will be briefly discussed in the Conclusions.
Finally,
we shall assume that $\phi$ and $\psi$ have arbitrary masses 
$m_\phi$ and $m_\psi$.

Given these assumptions, 
a straightforward calculation of the cross-section $\sigma_i$ for the process shown in  
Fig.~\ref{fig:diagrams}(a) then yields the result
\beq
 \sigma_i ~=~ \frac{g_\chi^2 g_\psi^2}{\pi (2E)^2}\, 
       \frac{ (E^2-m_i^2)^{1/2} \, (E^2 - m_\psi^2)^{3/2} }{(4 E^2 -m_\phi^2)^2}~, 
\eeq
where $E$ is the energy of each incoming $\chi$ particle in the center-of-mass frame.
Because the dark matter is assumed non-relativistic at the time of freeze-out, we may approximate
$E^2 \approx m_i^2 (1+v^2/4)$ where $v$ is the dark-matter relative velocity, whereupon we find that
\beq
  \sigma_i v ~=~ \frac{ g_\chi^2 g_\psi^2 }{128 \pi m_i^2} \, v^2 \, 
   \frac{(  1- {m_\psi^2/m_i^2})^{3/2}}
        {( 1- {m_\phi^2/4 m_i^2})^{2}}~. 
\eeq 
Recognizing that $\langle v^2 \rangle \sim T/m_i$ and that $T_i\approx m_i/20$ up to logarithmic
corrections, where $T_i$ is the freeze-out temperature of $\chi_i$, we then obtain the thermally averaged cross-section
\beq
  \langle \sigma_i v \rangle ~\sim~ \frac{ g_\chi^2 g_\psi^2 }{m_i^2} \, 
   \frac{ ( 1- {m_\psi^2/m_i^2})^{3/2}}
        {( 1- {m_\phi^2/4 m_i^2})^{2}}~, 
\label{finalcross}
\eeq 
where we are henceforth disregarding overall numerical factors.

For any such thermally averaged cross-section, the process of thermal freeze-out results in 
a present-day abundance given 
to leading order by~\cite{ZeldovichFreezeout,ChiuFreezeout,SteigmanFreezeout,ScherrerTurner} 
\begin{equation}
  \Omega_i ~\approx~ \OmegaDM 
    \left(\frac{\langle \sigma_i v \rangle}{1~\pb} \right)^{-1} 
    e^{-\Gamma_i t_{\rm now}}~,
\label{eq:OmegaDMLeadingOrder}
\end{equation}
where $\tnow\approx 13.8$~Gyr is the present age of the universe and where 
$\Gamma_i$ denotes 
the decay rate of the ensemble constituent $\chi_i$.  
Other than the $m_i$-dependence within $\sigma_i$ and $\Gamma_i$, 
this result is independent of $m_i$, up to logarithmic corrections.
This expression for $\Omega_i$ is predicated on the assumption that $\chi_i$ is its own antiparticle;
otherwise an additional overall factor of two would appear on the right side
of this equation.  
For simplicity, we shall henceforth utilize the
leading-order result in Eq.~(\ref{eq:OmegaDMLeadingOrder}).
We thus find that our resulting present-day 
cosmological abundance for this dark-matter component is given by
\beq
  \Omega_i ~\sim~ \frac {m_i^2}  { g_\chi^2 g_\psi^2 }
      ~ \frac{ ( 1- {m_\phi^2/4 m_i^2})^{2} }
         {( 1- {m_\psi^2/m_i^2})^{3/2}}
             ~ e^{-\Gamma_i t_{\rm now}}~.
\label{simplescaling}
\eeq

In the remainder of this paper, we shall focus on those
ensemble constituents $\chi_i$ for which $\Gamma_i \tnow \ll 1$ --- \ie,
those components whose decays have a negligible effect on $\Omega_i$.
We do this because these are precisely the ensemble components which survive
today and whose  abundances contribute to the present-day measurement of $\Omega_{\rm CDM}$. 
However, in this connection we remark that it is also possible to use the formalism
we develop in this paper
to study the abundances of those components whose lifetimes $\tau_i\equiv 1/\Gamma_i$ 
are significantly shorter than $t_{\rm now}$ and which have therefore already 
experienced significant decay prior to $t_{\rm now}$.
Indeed, for such components one natural approach would be to concentrate on the abundance $\Omega_i(t)$
of each component at $t=\tau_i$,
as we expect such abundances
to also obey the same types of inverse scaling relations that we expect for the abundances
of those components surviving today~\cite{DDM1}.   However, 
However, comparing abundances at different times 
    $\tau_i$ during cosmological history involves an 
    additional complication.  In particular, we must 
    to take into account the differing abundance 
    rescalings~\cite{DDM1} that arise due to the differences 
    between the decay timescales~$\tau_i$ of the 
    different ensemble constituents relative to 
    $t_{\rm now}$ --- timescales which may potentially 
    even extend into different cosmological epochs.

Within our result in Eq.~(\ref{simplescaling}),
it is natural to assume that $m_i \gg m_\psi$ for all $i$, as this is the condition that
underpins the process $\chibar_i\chi_i \to \phi \to \psibar\psi$ which eventually
induces the thermal freeze-out of $\chi_i$ and $\chibar_i$.
We shall also assume that 
$\Gamma_i \tnow \ll 1$, as described above.
Under such circumstances, we then find that the dark-matter abundance scales with $m_i$ as
\beq
  \Omega_i ~\sim~ {m_i^2}  \, \left( 1- \frac{m_\phi^2}{4 m_i^2}\right)^{2}~. 
\label{finalscaling}
\eeq

In principle, there is no specified relationship between $m_i$ and $m_\phi$.
However, we see from Eq.~(\ref{finalscaling}) that whether 
or not $\Omega_i$ increases or decreases with $m_i$ depends crucially on this relationship.
In particular, we have the opposite limiting cases:
\beqn
    m_i \gg m_\phi:& ~~\Omega_i \sim m_i^2~~ ~~& \Longrightarrow~ \gamma = +2 ~,\nonumber\\ 
    m_i \ll m_\phi:& ~~\Omega_i \sim m_i^{-2} ~~& \Longrightarrow~ \gamma = -2 ~.~~
\label{opplimits}
\eeqn
Clearly the first case with an extremely light mediator $\phi$
yields the canonical scaling behavior that we already discussed in connection to the WIMP miracle.
However, as promised, we see that the process of increasing the mediator mass $m_\phi$
and ultimately taking $m_\phi\gg m_i$ 
results in a flipping of the sign of the scaling exponent $\gamma$
from positive to negative values. 
Thus, in this regime, the cosmological abundances $\Omega_i$ resulting from thermal freeze-out
decrease with increasing mass $m_i$ --- precisely as desired.
Indeed, this result remains true for all ensemble constituents whose masses $m_i$ are significantly
smaller than the intermediary mass $m_\phi$.

For dark-matter ensembles whose constituent masses are capped at some maximum value $m_{\rm max}$, 
taking $m_\phi \gg m_{\rm max}$ ensures that our desired scaling relationship holds across the entire
dark-matter ensemble. 
Thus, in such cases, thermal freeze-out can indeed serve as a viable production mechanism within
the DDM framework.
However, in many theoretical constructions our resulting dark-matter ensemble contains
an infinite number of constituents whose masses grow without bound.   In such cases, our desired
scaling behavior holds only across that (lighter) portion of the ensemble for which $m_i\ll m_\phi$.
Indeed, as $m_i$ increases and becomes commensurate with $m_\phi$, 
other effects become relevant.  For example, for constituents with $m_i \sim m_\phi/2$,
resonance effects become important.  In this regime, provided that $\Gamma_\phi$ is not
too small, we find that $\Omega_i \propto \Gamma_\phi^2$, where $\Gamma_\phi$ is the total width of
the mediator.  (For an extremely small mediator width, $\Omega_i$ is sensitive to the
velocity distribution of $\chi_i$ and thus has a different parametric
dependence~\cite{NarrowWidth1,NarrowWidth2};
moreover, energy-dependent corrections to $\Gamma_\phi$ can also have an effect on $\Omega_i$, as discussed
in Ref.~\cite{Grz}.)
Finally, as
$m_i$ increases even further, the corresponding abundances $\Omega_i$  ultimately begin to increase.  Moreover,
for $m_i \geq m_\phi$, annihilation to a pair of on-shell mediators
becomes kinematically accessible.  Since the corresponding cross-section does not have the
same parametric dependence on $g_\chi$, $m_i$, and $m_\phi$ as 
in Eq.~(\ref{finalcross}), our scaling relation for the abundances in
Eq.~(\ref{simplescaling}) no longer holds in this regime.  

Within most DDM models it is usually the lighter dark-matter constituents 
which play the most significant roles within the resulting dark-sector phenomenology.
This is true for collider signatures~\cite{DDMLHC1,DDMLHC2}
as well as constraints coming from direct- and indirect-detection
experiments~\cite{DDMDD,DDMAMS,cosmicray1,cosmicray2}.  
Moreover, extremely heavy states within the ensemble may be expected 
to decay extremely rapidly in the early universe,
potentially prior to the epochs during which such decays could run afoul of standard cosmological constraints
and prior to the stage at which such states would experience thermal freeze-out.
Thus, for most practical concerns, our main focus is usually on those lighter components of the ensemble which
are most likely to survive to the present day and thus have the greatest phenomenological relevance.
Fortunately, our mechanism for flipping the scaling of the abundance spectrum   
applies precisely for those dark-matter constituents. 
These issues will be discussed further in Sect.~\ref{sec:DecayRateScaling}.

Thus, we conclude that 
thermal freeze-out can serve
as  a suitable abundance-production mechanism within the DDM framework. 
Indeed, within the annihilation channel we have considered here, we need only 
ensure that the mediator mass $m_\phi$ significantly exceeds $m_i$ over all relevant portions of the
DDM ensemble.   The mediator mass $m_\phi$ 
can then serve as a free parameter which may be
adjusted so as to render $\Omega_{\rm tot}\equiv \sum_i \Omega_i$ 
equal to $\Omega_{\rm CDM}$, as desired.

\section{Generating a spectrum of scaling exponents:  ~A more general study\label{general}}

Thus far, we have shown that we can flip the sign of the abundance scaling exponent $\gamma$ from $+2$ to
$-2$.   This then produces a negative scaling exponent, consistent with our original goal.
However, it is interesting for the purpose of considering many possible dark-sector phenomenologies
and for general model-building purposes to explore the full range of values of $\gamma$ which
may arise when our underlying annihilation process is varied.  
Moreover, even within the specific annihilation process we have considered,
it is interesting to study more general cases beyond that in which $m_\phi/m_i$ is 
taken to infinity.
Finally, as we shall see, there can also be final-state kinematic effects which 
we have thus far ignored but which might  
also potentially
affect the values of the scaling exponent $\gamma$.
We shall now undertake a general study of 
all of these possibilities.

Clearly, in order to obtain a variety of values of $\gamma$,
one might consider a corresponding variety of dark-matter annihilation processes
beyond that sketched in Fig.~\ref{fig:diagrams}.    
Indeed, there is almost no limit 
to the complexity of annihilation processes which might be considered. 
However, it is also interesting to remain
within the class of annihilation diagrams sharing the very natural topology
of that in Fig.~\ref{fig:diagrams}, but to consider alternative options for the spins of the internal
and external particles as well as alternative Lorentz structures for the couplings between
the dark and visible sectors.
We shall follow the latter course in this paper.

Towards this end, let us reconsider the annihilation diagram 
in Fig.~\ref{fig:diagrams}(a).
We shall again consider a toy DDM model in which the annihilation rate for each ensemble 
constituent $\chi_i$ in the early universe is dominated by this $s$-channel process, 
and we shall again not make any assumptions concerning 
the specific identities of $\chi_i$ and $\psi$ 
except that $\psi$ is presumed lighter than $\chi_i$ 
for all $i$.
We shall likewise not specify whether $\chi$ and $\psi$ are their own antiparticles.
However, we shall now allow $\chi_i$ and $\psi$ to be either complex scalars or spin-1/2 fermions.
Likewise, we shall allow $\phi$ to be either a spin-0 or spin-1 field.
Furthermore, we shall allow the couplings  
between the mediator 
and the dark and visible sectors 
to have a variety of Lorentz structures:  scalar (S), pseudoscalar (P), vector (V), or axial vector (A),
constrained only as appropriate for the particle spins involved.
In each case, we shall again assume that these couplings are the same for each ensemble constituent (and hence 
independent of the $i$ index),
and in each case we shall again consider only the leading (renormalizable or super-renormalizable) operators. 

The resulting possibilities are enumerated in Tables~\ref{Tab:X} and \ref{Tab:psi}.
Note that the cases with spinless mediators coupled to spinless dark or visible matter  
give rise to super-renormalizable interactions;   they thus depend on an arbitrary energy scale $\mu$.
Moreover, unlike all other cases, those involving a spin-1 mediator and spinless
dark or visible matter necessarily require derivative couplings.
Finally, for logical consistency and completeness,
we have included couplings involving the time-like components of vectorial interactions.
However, these operators
cannot couple to any external (initial or final) state regardless 
of the charge-conjugation, parity, or angular-momentum quantum numbers which that state might carry~\cite{Kumarfatia}. 
These couplings thus need not be considered further.

%============== BEGIN TABLES ===================================
\begin{table}[t]
\centering
\begin{tabular}{||c|c|c||c|c||}
  \hline
  \hline
  ~$\chi_i$~ & ~$\phi$~ & ~coupling~ & ~$\epsilon_\chi$~ & ~$r$~ \\
  \hline
  ~spin-0~ & ~spin-0~               & ~S:~~ $g_\chi\mu \chi^\ast \chi \phi$~ & 1 & 0 \\
  \hline
  ~spin-1/2~ & ~spin-0~             & ~S:~~ $g_\chi \chibar\chi \phi$~       & 0 & 1 \\
  ~spin-1/2~ & ~spin-0~             & ~P:~~ $g_\chi \chibar \gamma_5 \chi\phi$~  & 0 & 0 \\
  \hline
  ~spin-0~   & ~spin-1 (time)~      & ~V:~~ $g_\chi (\chi^\ast \partial_0 \chi) \phi^0$~ & --- & --- \\
  ~spin-0~   & ~spin-1 (spatial)~   & ~V:~~ $g_\chi (\chi^\ast \partial_i \chi) \phi^i$~ & 0 & 1 \\
  \hline
  ~spin-1/2~ & ~spin-1 (time)~      & ~V:~~ $g_\chi \chibar \gamma_0 \chi \phi^0$~       & --- & --- \\
  ~spin-1/2~ & ~spin-1 (spatial)~   & ~V:~~ $g_\chi \chibar \gamma_i \chi \phi^i$~       & 0 & 0 \\
  ~spin-1/2~ & ~spin-1 (time)~      & ~A:~~ $g_\chi \chibar \gamma_0 \gamma_5 \chi \phi^0$~ & 0 & 0 \\
  ~spin-1/2~ & ~spin-1 (spatial)~   & ~A:~~ $g_\chi \chibar \gamma_i \gamma_5\chi \phi^i$ ~ & 0 & 1 \\
   \hline
   \hline
\end{tabular}
\caption{~Values of the indices $\epsilon_\chi$ and $r$ which correspond to different  
  spins and coupling structures for the ensemble constituents $\chi_i$ and the 
  mediator $\phi$.
  \label{Tab:X}}
\vskip 0.25cm
\centering
\begin{tabular}{||c|c|c||c|c|c||}
  \hline
  \hline
  $~\phi$~ & ~$\psi$~ & coupling & ~$\epsilon_\psi$~ & ~$s$~ & ~$t$~ \\
  \hline
  ~spin-0~   & ~spin-0~           & ~S:~~  $g_\psi\mu \phi \psi^\ast \psi $~        & 1 & 0 & 0 \\
  \hline
  ~spin-0~ & ~spin-1/2~           & ~S:~~ $g_\psi \phi \psibar\psi $~        & 0 & 1 & 0 \\
  ~spin-0~ & ~spin-1/2~           & ~P:~~ $g_\psi \phi \psibar \gamma_5 \psi$~   & 0 & 0 & 0 \\
  \hline
  ~spin-1 (time)~   & ~spin-0     & ~V:~~ $g_\psi \phi^0 (\psi^\ast \partial_0 \psi) $~         & --- & --- & --- \\
  ~spin-1 (spatial)~   & ~spin-0  & ~V:~~ $g_\psi \phi^i (\psi^\ast \partial_i \psi) $~         & 0 & 1 & 0 \\
  \hline
  ~spin-1 (time)~ & ~spin-1/2~    & ~V:~~  $g_\psi \phi^0 \psibar \gamma_0 \psi $~         & --- & --- & --- \\
  ~spin-1 (spatial)~ & ~spin-1/2~ & ~V:~~  $g_\psi \phi^i \psibar\gamma_i \psi $~         & 0 & 0 & 0 \\
  ~spin-1 (time)~    & ~spin-1/2~ & ~A:~~ $g_\psi \phi^0 \psibar \gamma_0 \gamma_5 \psi $~  & 0 & 0 & 1 \\
  ~spin-1 (spatial)~ & ~spin-1/2~ & ~A:~~ $g_\psi \phi^i \psibar \gamma_i \gamma_5\psi $ ~   & 0 & 1 & 0 \\
   \hline
   \hline
\end{tabular}
\caption{~Values of the indices $\epsilon_\psi$, $s$, and $t$ which correspond to different 
  spins and coupling structures for the mediator $\phi$ and the  
  particle species $\psi$ into which the ensemble constituents annihilate.
  \label{Tab:psi}}
\end{table}
%============== BEGIN TABLES ===================================

Also shown in Tables~\ref{Tab:X} and \ref{Tab:psi} are the values of 
certain corresponding indices $(\epsilon_\chi,\epsilon_\psi,r,s,t)$.
The indices $\epsilon_\chi$ and $\epsilon_\psi$ indicate the overall power of the energy scale $\mu$
that is needed in the corresponding coupling.  (Equivalently, these indices are given by $4-d$ where
$d$ is the mass dimension of the corresponding Lagrangian operator.)
Likewise, we define $r=0$ if the mediator $\phi$ can couple to an initial state with total
angular momentum $L=0$, and $r=1$ if the mediator can only couple to an initial state
with $L=1$.  Similarly, we define $s=0$ if $\phi$ can couple to a final state with $L=0$,
and $s=1$ if $\phi$ can only couple to a final state with $L=1$.  Finally, we define $t=1$
if the coupling between $\overline{\psi}$, $\psi$, and $\phi$ is chirality-suppressed,
and $t=0$ otherwise.
Note that if $\phi$ is a spin-1 particle with pseudovector couplings to both
$\chi_i$ and $\psi$, contributions involving both the timelike and spacelike
components of $\phi$ must be included.  In cases in which $\psi$ is very light,
the space-like components yield the dominant contribution.  By contrast, when
$m_\psi \sim m_i /2$, the time-like component dominates.

In order for the process 
$\overline{\chi}_i \chi_i \rightarrow \phi \rightarrow \overline{\psi}\psi$ to
dominate the  annihilation rate for each $\chi_i$, the contribution to that rate from   
co-annihilation processes of the form 
$\overline{\chi}_i \chi_j \rightarrow \overline{\psi} \psi$ with $i\neq j$ 
must be suppressed.  This occurs naturally, for example, in scenarios in which
each of the $\chi_i$ is non-trivially charged under a different approximate symmetry.  In 
addition, the collective contribution to the annihilation rate 
from intra-ensemble annihilation processes of the form $\overline{\chi}_i \chi_i \rightarrow \overline{\chi}_j \chi_j$, 
in which heavier ensemble constituents annihilate into lighter ones, must likewise be 
suppressed.  This occurs naturally in scenarios in which 
$g_\chi \mu^{\epsilon_\chi} \ll g_\psi \mu^{\epsilon_\psi}$.  However, as we shall see, 
a suppression of this sort arises in a variety of other contexts as well.

Given these assumptions concerning the nature of the dominant dark-matter
annihilation processes, we can now calculate the corresponding cosmological abundances $\Omega_i$ that
emerge after thermal freeze-out.
For each spin/coupling combination in Tables~\ref{Tab:X} and \ref{Tab:psi}, the corresponding annihilation 
matrix elements $|\calM|^2$, summed over final states and averaged over initial states, scale as
\beqn
  |\calM|^2 &\sim& g_\chi^2 g_\psi^2 \, v^{2r} \, \left( {\mu\over m_i}\right)^{2 (\epsilon_\chi+\epsilon_\psi)} \nonumber\\
          && ~~\times~ \frac
            {  (1-m_\psi^2/m_i^2)^{s} }
            {  (1-m_\phi^2/4 m_i^2)^2 } \,
          \left( {m_\psi\over m_i}\right)^{2t}~,~~~
\eeqn
where the corresponding values of $\epsilon_\chi$, $\epsilon_\psi$, $r$, $s$, and $t$ are listed in 
Tables~\ref{Tab:X} and \ref{Tab:psi}.
The corresponding annihilation cross-sections then scale as
\beqn
  \sigma_i &\sim&  {g_\chi^2 g_\psi^2\over m_i^2}  \, v^{2r-1} \, \left( {\mu\over m_i}\right)^{2 (\epsilon_\chi+\epsilon_\psi)} \nonumber\\
         && ~~\times~
          \frac
            {  (1-m_\psi^2/m_i^2)^{s+1/2} }
            {  (1-m_\phi^2/4 m_i^2)^2 } \,
          \left( {m_\psi\over m_i}\right)^{2t}~.~~~
\eeqn
Calculating $\langle\sigma_i v\rangle$ from these results is not difficult.
As in Sect.~\ref{sec:ThermalFreezeout},
we focus on the regime in which each
$\chi_i$ freezes out at a temperature $T_i$ such that $x_i \equiv m_i/T_i \gg 3$.
In this regime, the velocity distribution for each $\chi_i$ is already non-relativistic,
with speed $v \ll 1$, by the time freeze-out occurs.  In this regime, the
cross-section is reasonably well approximated by retaining the leading non-vanishing
term in the series expansion
\begin{equation}
  \sigma_i v ~=~ a_i^{(0)} + a_i^{(1)} v^2 + a_i^{(2)} v^4 + \ldots 
  \label{eq:sigmavexpand}
\end{equation}
in the quantity $v^2$.  Recognizing that $\langle v^2 \rangle \sim T/m_i$,
we find that the corresponding thermal average at temperatures
$T \sim T_i$ is given by~\cite{SrednickiSigmavExpand}
\begin{equation}
  \langle \sigma_i v \rangle ~\approx~ a_i^{(0)} + 
    \frac{3}{2}a_i^{(1)} x^{-1}_i + \frac{15}{8} a_i^{(2)} x^{-2}_i + \ldots~.
  \label{eq:sigmavthermalexpand}
\end{equation}
Thus, in cases in which the annihilation is $s$-wave and the constant term
$a_i^{(0)}$ dominates, the thermal average $\langle \sigma_i v \rangle$
scales with $m_i$ and $g_i$ across the ensemble in exactly the same way as
$\sigma_i v$ itself.  Moreover, since the ratio $x_i$ depends only logarithmically
on $m_i$ and $g_i$ (due to the implicit dependence of $T_i$ on $m_i$), we find that
even in cases in which $a_i^{(0)}=0$ and the annihilation is $p$-wave,
$\langle \sigma_i v \rangle$ still scales with these parameters in approximately the
same way as $\sigma_i v$, up to logarithmic corrections.

Given these results, we then find from Eq.~(\ref{eq:OmegaDMLeadingOrder})
that our final abundances $\Omega_i$ scale across the ensemble as
\beq
  \Omega_i ~\sim~ {m_i^2 \over g_\chi^2 g_\psi^2} \, m_i^{2(\epsilon_\chi+\epsilon_\psi + t)} \, 
          \frac
            {  (1-m_\phi^2/4 m_i^2)^2 } 
            {  (1-m_\psi^2/m_i^2)^{s+1/2} } ~,~
\label{finalresult}
\eeq
where we continue to assume $\Gamma_i t_{\rm now}\ll 1$
and thereby ignore the effects of particle decays, focusing instead
on the original abundance produced by thermal freeze-out.
In this expression, it is easy to understand the origins of each factor:
\begin{itemize}
\item  the leading factor of $m_i^2$ is nothing but the canonical contribution that
           already appeared in Eqs.~(\ref{WIMPmiracle}) and (\ref{tradresult});
\item  the second factor $m_i^{2(\epsilon_\chi+\epsilon_\psi+t)}$ reflects the possibility of 
       super-renormalizable couplings in Tables~\ref{Tab:X} and \ref{Tab:psi},
       and also reflects the possibility of a chirality-suppressed 
       coupling between the mediator and the visible sector;
\item  the third factor $(1-m_\phi^2/4m_i^2)^2$ reflects the
       contribution from the mediator; and
\item  the final factor $(1-m_\psi^2/m_i^2)^{-s-1/2}$ reflects final-state
       kinematic effects.
\end{itemize}

If we consider only the first and third factors, we reproduce the result
in Eq.~(\ref{finalscaling}).   
Indeed, we now see that it is legitimate to consider only the first and third factors 
in those cases for which $\epsilon_\chi=\epsilon_\psi=t=0$ (thereby eliminating the second factor)
and for which $m_i\gg m_\psi$ for all $i$ (thereby eliminating the fourth factor).
In such cases, we then reproduce our prior results in Eq.~(\ref{opplimits}), with $\gamma$ flipping from $+2$
to $-2$ when the mediator is taken from extremely light to extremely heavy.

It is now apparent, however, that there are additional effects which can come into play.
First, there is the contribution from the second factor.   In general, the contribution
from this factor {\it increases}\/ the value of the scaling exponent by an amount
\beq
  \Delta \gamma ~\equiv ~  2 (\epsilon_\chi + \epsilon_\psi+t)~.
\label{extraconts}
\eeq
It is immediately apparent from the various self-consistent coupling and mediator combinations 
in Tables~\ref{Tab:X} and \ref{Tab:psi} that the only allowed values for $\Delta \gamma$
are $0$, $2$, and $4$.
We thus see that the possibility of super-renormalizable
couplings and chirality-suppressed mediator/visible-sector couplings
tends to drive $\gamma$ towards even more positive values.
Indeed, for those combinations with $\Delta \gamma=4$, 
this effect completely cancels the effect from 
integrating out the mediator
(\ie, the effect from taking $m_\phi\gg m_i$ for all $i$), 
restoring our positive traditional scaling exponent $\gamma=2$. 
However, the case with $\Delta \gamma=2$ leaves
us with $\gamma=0$, producing
cosmological abundances which are largely independent of the constituent masses $m_i$ to within the
leading approximations we have been making.
In such circumstances, it is then the {\it subleading}\/ contributions to thermal freeze-out
(coming perhaps from non-dominant annihilation channels and subleading contributions to the 
thermal averaging process, {\it etc.}\/) 
which dictate the overall sign of the scaling exponent $\gamma$.

Finally, we consider the 
final-state kinematic effects coming from the fourth factor.
When $m_i\gg m_\psi$ for all $i$, 
these effects are essentially independent of $m_i$ and thus do not alter 
the value of $\gamma$.
Otherwise, when we merely have 
$m_{\rm min}\gsim  m_\psi$ where $m_{\rm min}$ is the minimum of the $m_i$, this fourth
factor enhances the cosmological 
abundances $\Omega_i$ but does so {\it increasingly weakly}\/ as a function of 
the constituent mass $m_i$.
In other words, this factor is greater than one but decreases as a function of $m_i$.
This then tends to provide a negative (although $m_i$-dependent) contribution to $\gamma$
which can again pull the overall value of the scaling exponent $\gamma$ towards negative values.

We thus see that the question of 
whether the resulting values of $\gamma$ are positive or negative depends on the balancing
between a number of factors governing the annihilation process.
The canonical factor gives a contribution $\gamma= 2$,
and the coupling and chirality factors make a potential additional positive contribution given  
in Eq.~(\ref{extraconts}).
By contrast, integrating out the mediator $\phi$ tends to decrease the value of $\gamma$ by four units,
as we have seen in Sect.~\ref{sec:ThermalFreezeout},
and this is then further decreased by the final-state kinematic effects. 

Thus far, we have treated $\gamma$ as if this scaling exponent were constant across the entire ensemble.
In other words, we have implicitly assumed that $\Omega_i$ exhibits a pure power-law dependence on $m_i$.
This is certainly true for the contribution from the canonical first factor in Eq.~(\ref{finalresult}),
and true even for the extra contribution in Eq.~(\ref{extraconts})
coming from the second factor.
It is also true for the third factor as long as we consider the mediator $\phi$
to be  either extremely light ($m_\phi\ll m_i$ for all $i$) or extremely heavy ($m_\phi\gg m_i$ for all $i$),
and it is trivially true even for the fourth
factor as long as we consider our final-state 
particles to also be extremely light, with $m_\psi \ll m_i$ for all $i$. 
However, 
the scaling-exponent contributions from the third and fourth factors 
in Eq.~(\ref{finalresult}) are generally $m_i$-dependent, which means that our total scaling exponent
$\gamma$ will also be $m_i$-dependent.
Indeed, recognizing this fact is critical if we wish to extend our analysis beyond the limiting approximations
outlined above.

Fortunately, it is not difficult to obtain the corresponding results for 
these mass-dependent scaling exponents $\gamma(m)$.
In general, we have seen from Eq.~(\ref{finalresult}) that
$\Omega(m)$ can be viewed as a continuous function which varies with the mass scale $m$ 
within the allowed range $m_\psi < m < m_\phi/2$
according to
\beq
  \Omega(m) ~\sim~ {m^2 \over g_\chi^2 g_\psi^2} \, m^{2(\epsilon_\chi+\epsilon_\psi + t)} \, 
          \frac
            {  (1-m_\phi^2/4 m^2)^2 } 
            {  (1-m_\psi^2/m^2)^{s+1/2} } ~.
\label{finalresult2}
\eeq
Given this, we
can define our effective scaling exponent at any value of $m$ 
via the relation $\Omega(m)\sim m^{\gamma}$,  or equivalently
\beq
     \gamma(m) ~\equiv~ { d\ln \Omega(m)\over d\ln m} ~=~ {m\over \Omega(m)} {d \Omega(m)\over dm}~.
\eeq
For the abundance in Eq.~(\ref{finalresult2}), we then find  
\beq
 \gamma(m) ~=~ 
           2 \, + \, \Delta\gamma  \, +\,  {1\over m^2/m_\phi^2-1/4} \, +  \, {2s+1 \over 1-m^2/m_\psi^2 }~,
\label{gammaresult}
\eeq
where $\Delta \gamma$ is given in Eq.~(\ref{extraconts}).
Indeed, the separate terms in Eq.~(\ref{gammaresult}) are the contributions
from the corresponding factors in Eq.~(\ref{finalresult2}).

\begin{figure}[t]
\includegraphics[keepaspectratio, width=0.42\textwidth]{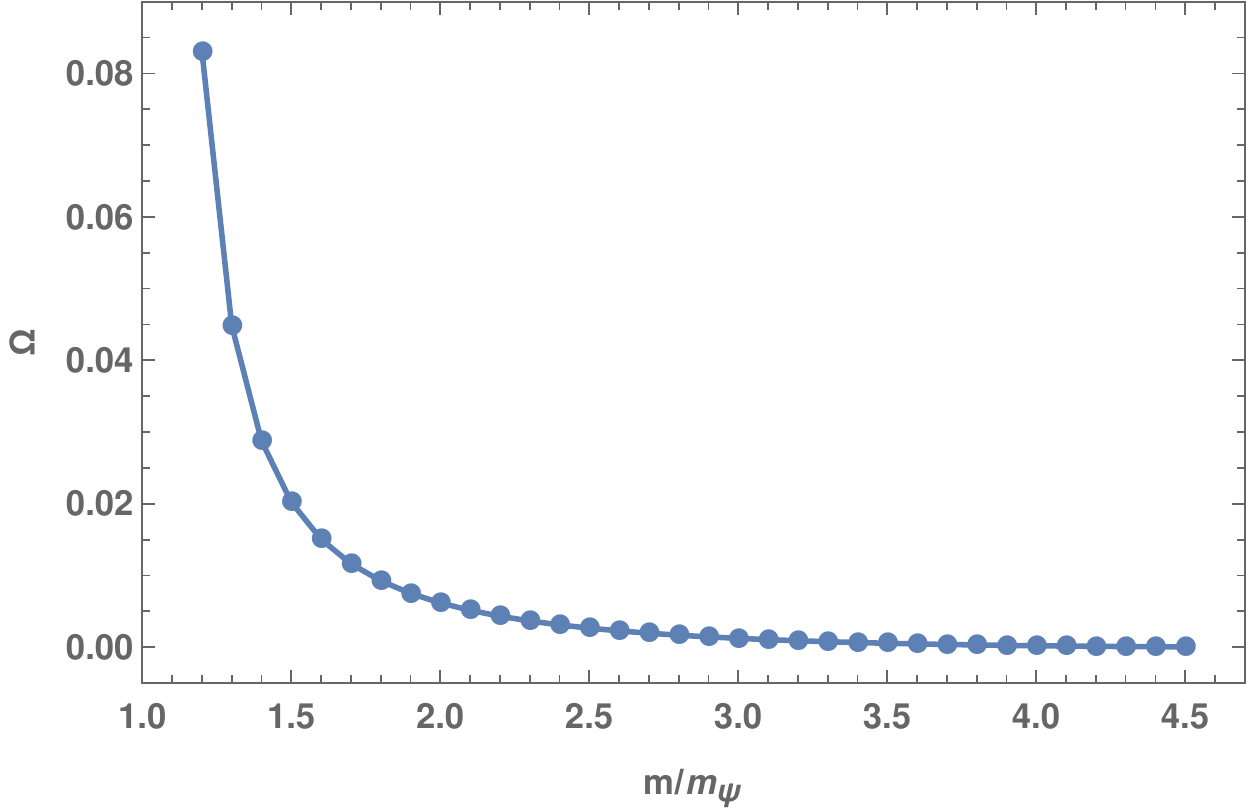}
\includegraphics[keepaspectratio, width=0.42\textwidth]{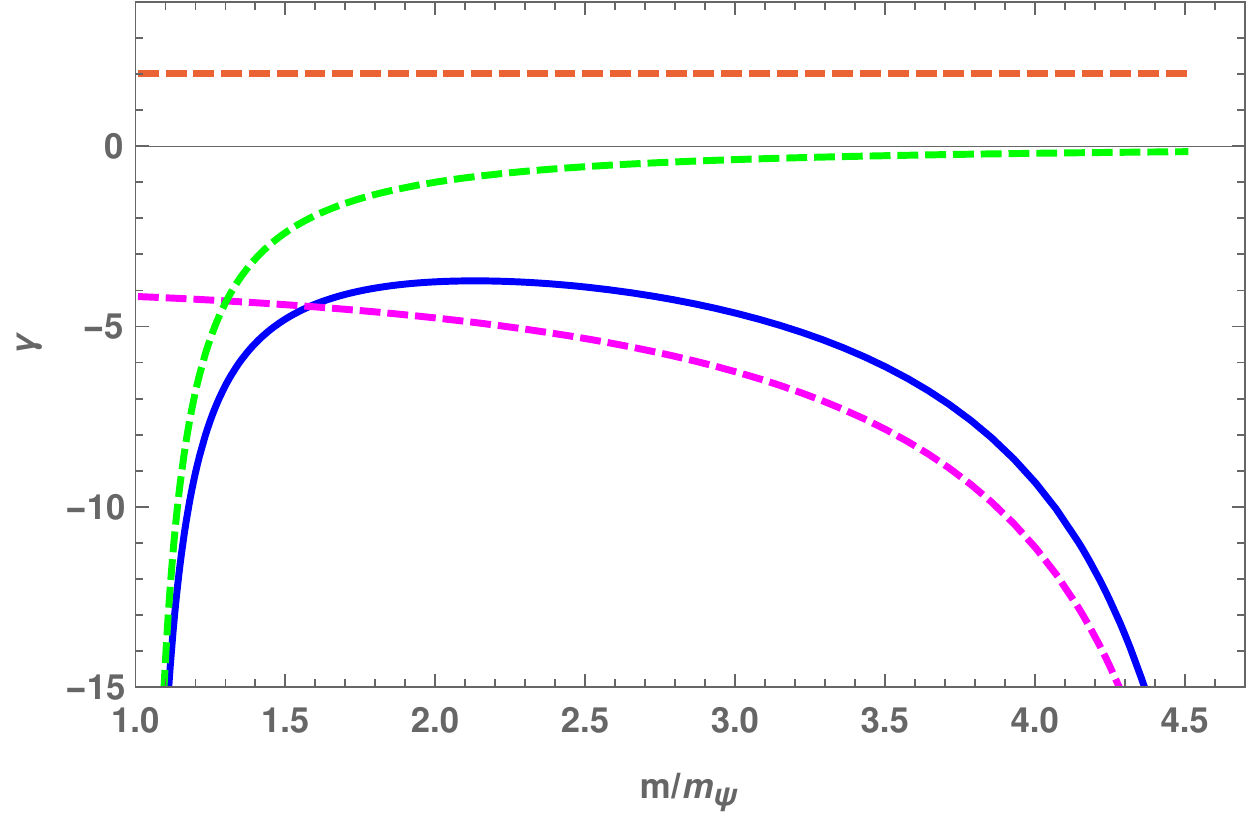}
\caption{Abundances $\Omega_i$ (upper panel) and contributions to the corresponding
scaling exponent $\gamma$ (lower panel) for $\epsilon_\chi=\epsilon_\psi=r=t=0$ and $s=1$.
We have taken $m_\phi=10 m_\psi$ and assumed a mass spectrum $m_i/m_\psi=\lbrace 1.2, 1.3, 1.4, ..., 4.5\rbrace$.
The corresponding values 
of $\Omega_i$ have been normalized so that $\sum_i \Omega_i=\Omega_{\rm CDM}\approx 0.26$.
We see from the upper panel that $\Omega_i$ indeed falls as a function of $m_i$, as desired, even though these
abundances arise from thermal freeze-out.  
The solid blue curve in the lower panel indicates the corresponding scaling exponent $\gamma(m)$, which 
is negative throughout the appropriate dark-matter mass range.   This curve receives additive contributions 
from the canonical contribution $\gamma=+2$ (red dashed line),  the effects of the heavy mediator $\phi$ (magenta dashed curve),
and the effects of final-state kinematics (green dashed curve).   
Given that the canonical contribution alone leads to $\gamma= +2$,
we see that the contributions from the latter two effects 
combine to pull this result into the $\gamma<0$ range and also to introduce a non-trivial
mass-dependence for $\gamma(m)$.}
\label{fig:massspectrum}
\end{figure}

In Fig.~\ref{fig:massspectrum}
we have plotted the results for $\Omega(m)$ and $\gamma(m)$ as functions of $m$
over the mass range $m_\psi < m < m_\phi/2$.
For these plots we have taken
$m_\phi=10 m_\psi$.
Within this mass range, 
we have also chosen a discrete 
mass spectrum $m_i/m_\psi=\lbrace 1.2, 1.3, 1.4, ..., 4.5\rbrace$,
and we have normalized the corresponding abundances $\Omega_i$ [and thus the 
overall abundance curve $\Omega(m)$ on which these abundances lie] so that
$\Omega_{\rm tot}\equiv \sum_i \Omega_i=\Omega_{\rm CDM}\approx 0.26$.
We have also taken our underlying annihilation process to 
have $\epsilon_\chi=\epsilon_\psi=r=t=0$ and $s=1$. 
Note that our choice of a particular discrete constituent mass spectrum $\lbrace m_i\rbrace$ 
populating the allowed mass range 
$m_\psi < m < m_\phi/2$
allows us to 
normalize our total cosmological abundance curve $\Omega(m)$ 
and thereby determine a
particular partitioning of $\Omega_{\rm CDM}$ across the different contributions $\Omega_i$.
Our choice of the discrete mass spectrum $\lbrace m_i\rbrace$
along these curves 
otherwise plays no essential role in fixing the behavior of $\Omega(m)$ as a function of $m$.

We see from the upper panel of Fig.~\ref{fig:massspectrum} that the ensemble abundances $\Omega_i$ fall 
as a function of the constituent masses $m_i$, as desired, even though these
abundances arise from thermal freeze-out.  This then verifies explicitly 
that thermal freeze-out can yield cosmological abundances which decrease, rather than increase, as a function of
the mass of the individual dark-matter constituents. 
Indeed, this behavior is smooth and resembles the behavior that has 
been observed for other non-thermal abundance-production mechanisms.

%---------------------BEGIN FIGURE-----------------------%
%----------------------END FIGURE------------------------%

The lower panel of Fig.~\ref{fig:massspectrum} illustrates the corresponding behavior of the
scaling exponent function $\gamma(m)$ in Eq.~(\ref{gammaresult}). 
The solid blue curve in the lower panel indicates the total scaling exponent $\gamma(m)$, which 
is negative throughout the appropriate dark-matter mass range.   
Indeed, this curve receives additive contributions from
the canonical result $\gamma=+2$ (red dashed line),  
the effects of the heavy mediator $\phi$ (magenta dashed curve),
and the effects of final-state kinematics (green dashed curve).   These are respectively the first, third, and fourth
terms in Eq.~(\ref{gammaresult}).
As already noted, we see that the contributions from the latter two effects 
combine to overwhelm the canonical contribution $\gamma= +2$ and pull the resulting scaling exponent into the 
$\gamma<0$ range.  They also introduce a non-trivial mass-dependence for $\gamma$.  

In the limit in which $m_\phi\to\infty$ (so that the mediator is fully
integrated out of the theory), the magenta curve starts at $\gamma= -4$ and remains essentially flat.  
Likewise, within mass regions for which $m\gg m_\psi$, the final-state kinematic effects disappear and the green curve
also becomes essentially flat at $\gamma=0$.
Thus, in these limits, we see that our canonical contribution $\gamma=+2$ 
is uniformly pulled down by the mediator effects to $\gamma = -2$,
as discussed in Sect.~\ref{sec:ThermalFreezeout}.
However, we now see that 
for masses $m$ which are not that far below $m_\phi/2$ 
or not that far above $m_\psi$
(\ie, for masses at the lighter and heavier ends of the allowed mass range),
the net effects of the mediator
and the final-state kinematics
are to bend the $\gamma(m)$ blue curve 
further below $\gamma = -2$.
This enhances the rates at which the corresponding abundances fall as functions of the mass.
Indeed, with both effects together, we find that the maximum value of $\gamma$ plotted in 
the lower panel of Fig.~\ref{fig:massspectrum} is approximately $-4$
rather than the value $\gamma= -2$ that would have existed without these effects.
Thus even the behavior of the central portion of the dark-matter ensemble 
is altered by these effects.

\begin{figure}[t]
\includegraphics[keepaspectratio, width=0.42\textwidth]{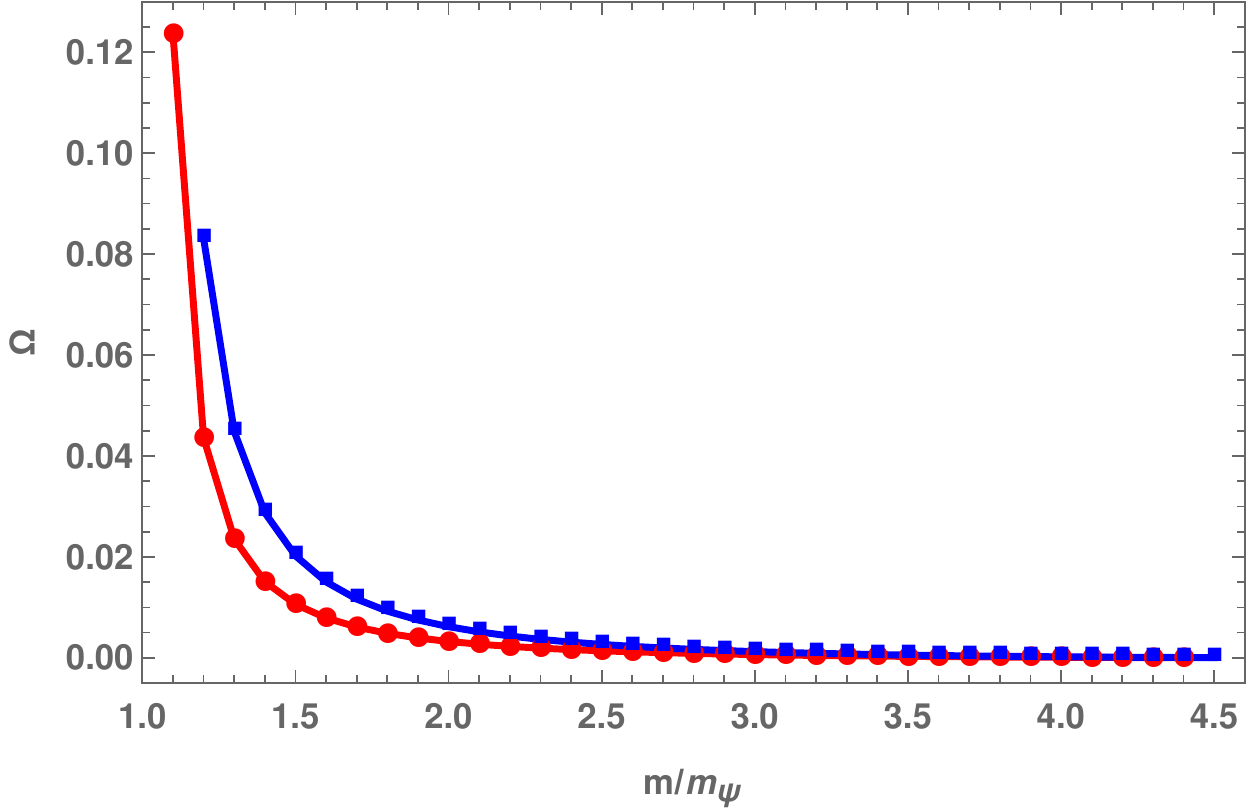}
\caption{ Abundances $\Omega_i$ 
with $s=1$, $\epsilon_\chi=\epsilon_\psi=r=t=0$, and $m_\phi/m_\psi=10$. 
The blue curve is the same as in the top panel of Fig.~\ref{fig:massspectrum},
corresponding to the discrete constituent mass spectrum
$m_i/m_\psi=\lbrace 1.2, 1.3, 1.4, ..., 4.5\rbrace$,
while the red curve corresponds to the same mass spectrum shifted downward by $\Delta m_i/m_\psi = -0.1$.
In each case the corresponding values 
of $\Omega_i$ have been normalized so that $\sum_i \Omega_i=\Omega_{\rm CDM}\approx 0.26$.
As a result of the falling behavior of $\Omega_i$ as a function of mass $m_i$, 
we see that the downward shift of our spectrum induces a renormalization
of all of the states and alters the magnitudes of the relative contributions 
to $\Omega_{\rm CDM}$, shifting the abundance distribution of the resulting DDM ensemble from 
one with $\eta\approx 0.65$ (blue) 
to one with $\eta \approx 0.52$ (red).  }
\label{fig:massspectrum_rel}
\end{figure}

It is important to note that while the specific choice of the discrete constituent mass spectrum $\lbrace m_i\rbrace$
within our overall allowed mass range has no effect on the behavior of the scaling exponents $\gamma(m)$,
this choice can nevertheless significantly affect the relative partitioning
of the total dark-matter abundance $\Omega_{\rm CDM}$ across the ensemble.
For example, in Fig.~\ref{fig:massspectrum_rel}
we have plotted the abundances for two different mass spectra:  that already plotted in the top
panel of Fig.~\ref{fig:massspectrum} (blue), and the spectrum that results by shifting the
mass of each component downward by $\Delta m_i / m_\psi = -0.1$ (red).
In each case, we have normalized the corresponding abundances $\Omega_i$ [and thus the corresponding
general $\Omega(m)$ curve] so as to hold
the total abundance $\Omega_{\rm tot}\equiv \sum_i \Omega_i = \Omega_{\rm CDM}\approx 0.26$ fixed.
Two observations are immediately apparent.
First, the blue curve is the same as the red curve
except for an overall multiplicative factor (which in this case is approximately $1.905$).
Second, however, the magnitudes of the relative constituent contributions $\Omega_i$
to the total abundance $\Omega_{\rm tot}$ are non-trivially altered due to the change in discrete masses.
One useful way to characterize the abundance distribution of a given DDM ensemble is
through the so-called ``tower fraction'' $\eta$, defined as~\cite{DDM1,DDM2}
\beq
      \eta ~\equiv~ 1 - { \Omega_{\rm max}\over \sum_i \Omega_i} ~~~~~ {\rm where}
              ~~ \Omega_{\rm max} \equiv {\rm max}_i \, \Omega_i~.
\eeq
Note that $0\leq \eta < 1$.  
In general, the value of $\eta$ indicates how much of the total abundance of the ensemble is carried by those states
which are {\it not}\/ the dominant one.  Thus smaller values of $\eta$
correspond to the more traditional ensembles in which only one or a few components carry the bulk 
of $\Omega_{\rm CDM}$,
while larger values of $\eta$ correspond to more DDM-like ensembles in which the abundance is more generally
distributed across the ensemble.
We see from the abundances plotted in Fig.~\ref{fig:massspectrum_rel} that the blue curve corresponds to 
approximately $\eta\approx 0.65$, while the red curve corresponds to approximately $\eta\approx 0.52$.
Thus, for abundances $\Omega_i$ which fall as a function of mass, shifting the spectrum of discrete constituent masses towards higher masses tends to increase the value of $\eta$ and thereby enhance the DDM-like nature of the corresponding ensemble.

%---------------------BEGIN FIGURE-----------------------%
%---------------------END FIGURE-----------------------%

Thus far we have only considered the case with 
$\epsilon_\chi=\epsilon_\psi=t=0$.   Thus $\Delta \gamma=0$, and the second factor in Eq.~(\ref{gammaresult}) has no effect.
Likewise, we have only considered the case with $m_\phi/m_\psi=10$, which is a relatively small mass hierarchy.
In principle --- and for phenomenological purposes --- we are interested in much larger values of this ratio.
We therefore turn to examine how the basic picture outlined above changes for $\Delta \gamma >0$ and for
larger values of $m_\phi/m_\psi$.

It is readily apparent that choosing spin and coupling structures in Tables~\ref{Tab:X} and \ref{Tab:psi} with non-zero
values of $\lbrace \epsilon_\chi,\epsilon_\psi, t\rbrace$ only serves to shift the $\gamma$-curves in Fig.~\ref{fig:massspectrum}
uniformly upwards by an amount $\Delta\gamma$.
For $\Delta \gamma = 2$, it is clear that the 
blue curve remains completely within the $\gamma< 0$ range.
Thus, for $\Delta \gamma =2 $ 
and for the value of $m_\phi/m_\psi=10$ chosen for the plots in Fig.~\ref{fig:massspectrum}, 
the abundances $\Omega_i$ continue to scale inversely with the dark-matter masses $m_i$ throughout the ensemble.
However, when shifted by $\Delta \gamma= +4$ --- 
as occurs when the couplings of the mediator to the dark and visible sectors are
both super-renormalizable, with $\epsilon_\chi=\epsilon_\psi=1$ ---
the blue curve 
within the lower panel of Fig.~\ref{fig:massspectrum} actually exceeds zero 
within the approximate region $1.3\lsim m/m_\psi\lsim 2.7$.
This means that the corresponding cosmological abundances $\Omega_i$ fall as a function of $m_i$ for
$m_i/m_\psi \lsim 1.3$ and then rise for 
$1.3\lsim m_i/m_\psi\lsim 2.7$ before ultimately falling 
again for 
$m_i/m_\psi \gsim 2.7$.

%---------------------BEGIN FIGURE-----------------------%
\begin{figure}[t]
\includegraphics[keepaspectratio, width=0.42\textwidth]{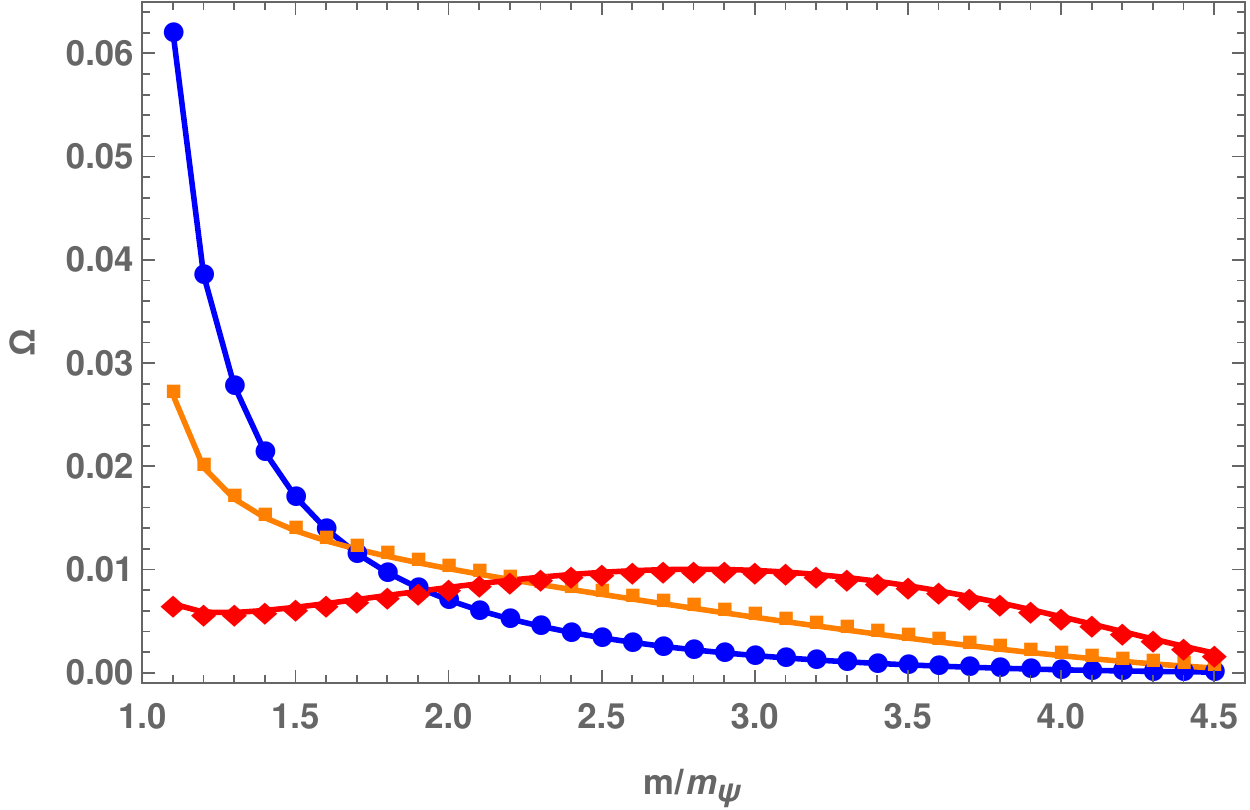}
\includegraphics[keepaspectratio, width=0.42\textwidth]{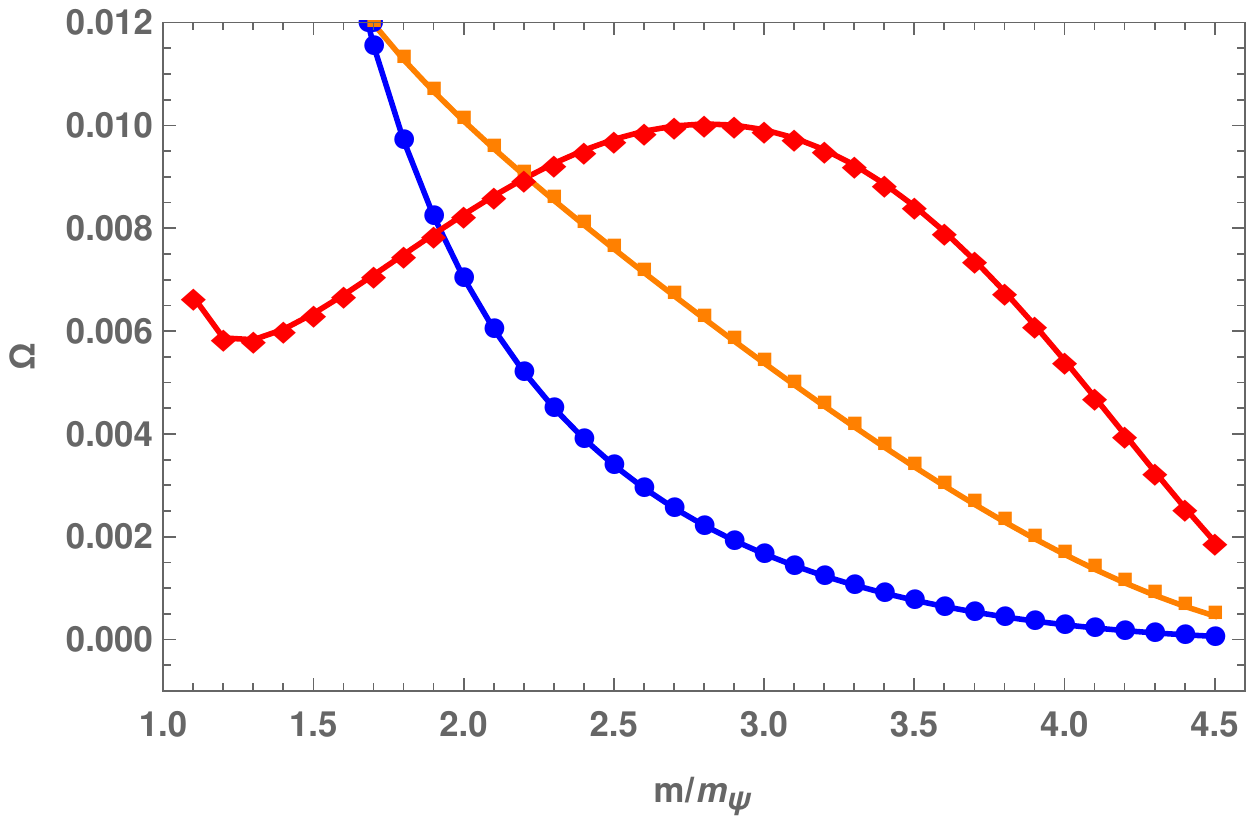}
\caption{Upper panel:   same as the upper panel of Fig.~\ref{fig:massspectrum} except with $s=0$, 
plotted for different values of 
$\Delta \gamma$.  The cases with 
$\Delta \gamma =0,2,4$ correspond to the blue, orange, and red curves, respectively.  
As in the upper panel of Fig.~\ref{fig:massspectrum}, each curve is individually normalized
so that the total abundance $\sum_i \Omega_i$ is fixed at $\Omega_{\rm CDM}\approx 0.26$.   Lower panel:  a zoom-in of the small-abundance portion of the upper panel.
We see that the abundance curves are monotonically decreasing as a function of $m$ for $\Delta \gamma = 0$
and $\Delta \gamma=2$,
while for  
$\Delta \gamma=4$ there is a region $1.3\lsim m/m_\psi\lsim 2.7$
over which the abundances increase as a function of $m$ before decreasing again.  }
\label{othercases}
\end{figure}
%----------------------END FIGURE------------------------%

%---------------------BEGIN FIGURE-----------------------%
\begin{figure}[t]
\includegraphics[keepaspectratio, width=0.42\textwidth]{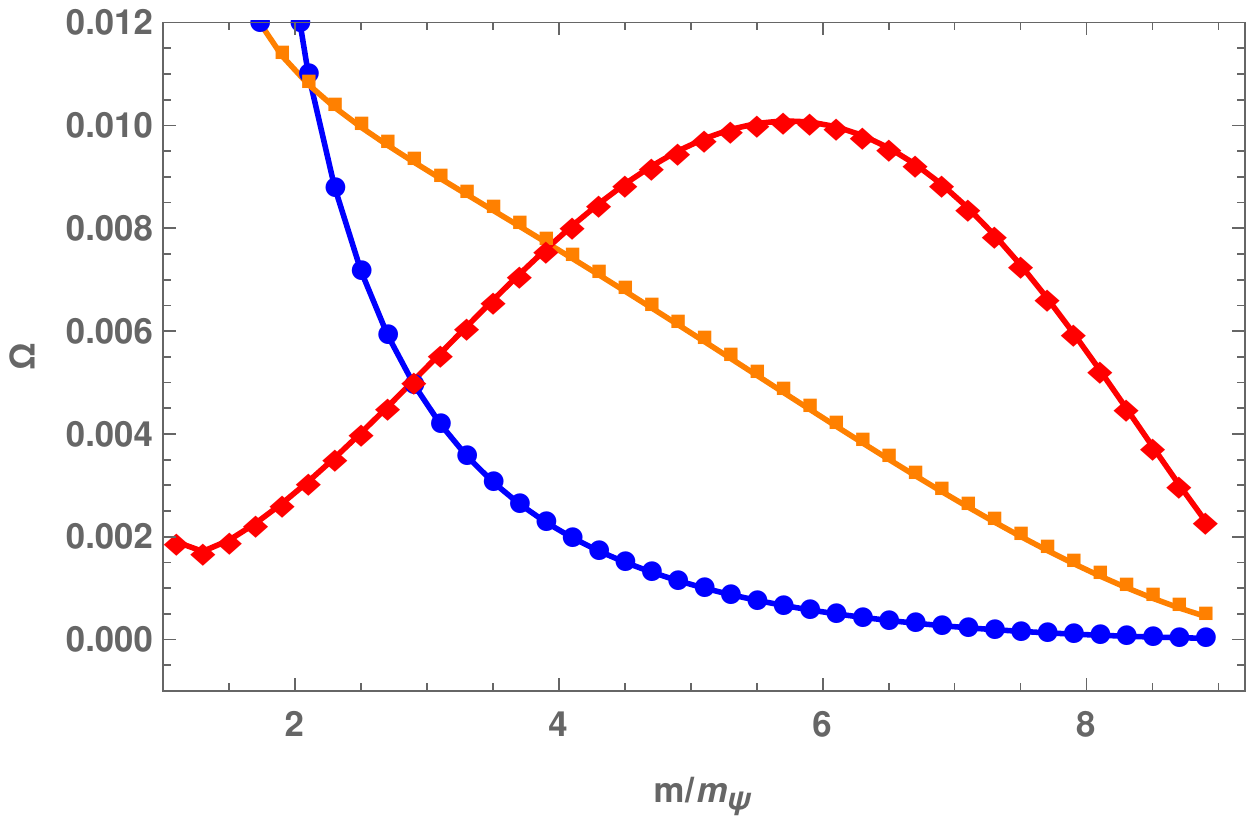}
\includegraphics[keepaspectratio, width=0.42\textwidth]{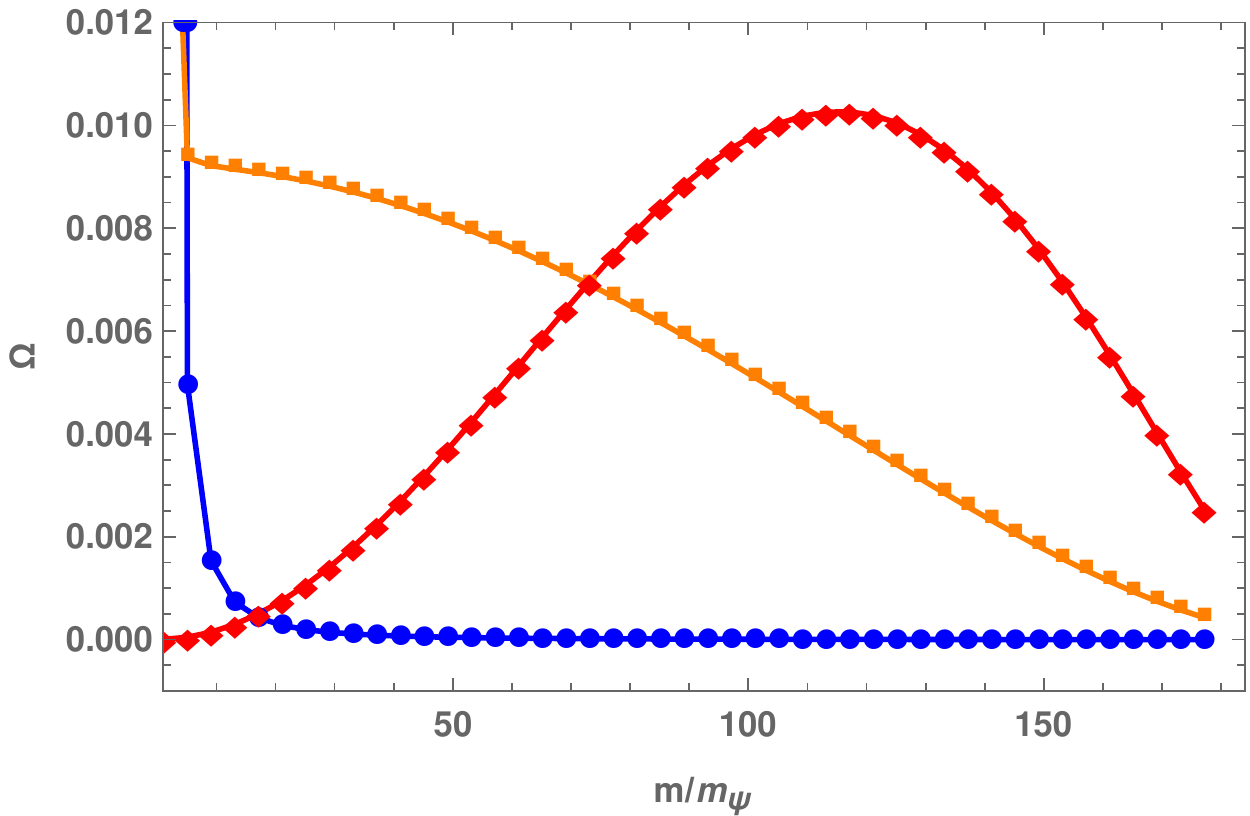}
\caption{ Same as lower panel of Fig.~\ref{othercases}, except with $m_\phi/m_\psi$ 
now increased to $m_\phi/m_\psi=20$ (upper panel)
and $m_\phi/m_\psi=400$ (lower panel), with our discrete constituent mass spectrum rescaled accordingly.
As $m_\phi/m_\psi$ increases,
the $\Delta \gamma=4$ curves (red) 
exhibit  increasingly pronounced local minima and maxima 
while the $\Delta \gamma =0,2$ curves (blue and orange)
remain monotonically decreasing.
The behavior of the curves shown in the lower panel
essentially illustrates the behavior that emerges in the asymptotic $m_\phi/m_\psi\to\infty$ limit.
Thus we see that for all values of $\Delta \gamma$ and $m_\phi/m_\psi$ there
exist relatively large (and occasionally unconstrained) regions 
of $m$ over which the corresponding cosmological abundances $\Omega$ fall as a function of $m$.} 
\label{othercases2}
\end{figure}
%----------------------END FIGURE------------------------%

The behavior of the cosmological abundance $\Omega(m)$ as a function of $m$
is shown in Fig.~\ref{othercases} for $\Delta\gamma =0$, $2$, and $4$.
The upper panel shows the full range of cosmological abundances realized
in these scenarios, while the lower panel shows the lower region of the upper
panel in more detail.
For these plots, we have taken $m_\phi/m_\psi=10$ and $s=0$.
We have also assumed the 
discrete mass spectrum $m_i/m_\psi= 1.1, 1.2, ..., 4.5$ and plotted
the corresponding values of $\Omega_i$.
In each case, an overall normalization has been chosen
so that $\Omega_{\rm tot}=\Omega_{\rm CDM}\approx 0.26$.
The blue curve in the upper panel (representing the $\Delta \gamma=0$ case)
can be compared with the red curve in Fig.~\ref{fig:massspectrum_rel}
in order to discern the effects of taking $s=0$ rather than $s=1$.
However, as expected, we now see that increasing $\Delta \gamma$ has
the net effect of decreasing the rate at which the
corresponding abundances $\Omega_i$ fall as functions of $m$.
Indeed, increasing $\Delta\gamma$ all the way to $4$
even manages to induce a localized mass region in which the abundances $\Omega_i$ actually {\it increase}\/ 
as a function of $m$.
Thus, for $\Delta\gamma=4$, we see that the effects from both
a heavy mediator and final-state kinematics 
have conspired to produce not only a {\it non-monotonic}\/ $\Omega(m)$ function, but with it
also a {\it local minimum}\/ for $\Omega(m)$,
as shown in Fig.~\ref{othercases}.
This could thereby give rise 
to a potentially interesting new phenomenology.
Indeed, in such cases we see that thermal freeze-out has
effectively selected a particular mass scale for special treatment,
endowing the corresponding member of the dark-matter ensemble 
with a small, extra bit of cosmological invisibility
as compared with its immediate lighter and heavier neighbors.

Finally, we consider the behavior that emerges for larger hierarchies $m_\phi/m_\psi$.
It is important to stress that this is {\it not}\/ the same as 
integrating out the mediator $\phi$, since we are still considering all possible values of $m/m_\phi< 1/2$ without
requiring $m \ll m_\phi$. 
In order to make meaningful comparisons with different rescaled values of $m_\phi/m_\psi$, we 
simultaneously 
rescale the mass differences across our assumed discrete mass spectrum.
In other words, the discrete mass spectrum
$(m_k-m_0)/m_\psi= 0.1 k$ 
with $m_0\equiv 1.1 m_\psi$
that we previously took 
for \hbox{$m_\phi /m_\psi=10$}
will now be taken as
$(m_k-m_0)/m_\psi= 0.01 (m_\phi/m_\psi) k$
for any value of $m_\phi/m_\psi$.
This reduces to the original mass spectrum for
$m_\phi /m_\psi=10$
but otherwise scales so as to similarly fill the allowed range $m_\psi \leq m \leq m_\phi/2$ 
while keeping the lightest component
anchored at $m_0 =1.1 m_\psi$.

The resulting cosmological abundances are shown in Fig.~\ref{othercases2} for $m_\phi/m_\psi=20$ (upper panel) and 
$m_\phi/m_\psi=400$ (lower panel).
As in Fig.~\ref{othercases},
we have once again taken $s=0$.
In general, for $\Delta \gamma=4$, we see that increasing the value of $m_\phi/m_\psi$ tends to 
enhance the non-monotonicity of the
abundance $\Omega(m)$ as a function of $m$ that we have already observed in
Fig.~\ref{othercases}.
By contrast, the cases with
$\Delta\gamma \leq 2$ remain
completely monotonic.

This behavior survives even as $m_\phi/m_\psi\to \infty$.
Indeed, for $m_\phi/m_\psi \to \infty$, we find that 
the cosmological abundances monotonically decrease as a function of $m$ for $\Delta \gamma =0,2$.
By contrast, for any $\Delta \gamma >2$, we find that
these abundances $\Omega(m)$  monotonically decrease as a function of $m$ within only two disconnected regions:
a very small region at low masses near $m_\psi$ given by 
\beq
   { m \over m_\psi}  ~<~ \sqrt{1 + {2s+1\over \Delta\gamma -2}}~,
\label{lowregime}
\eeq
and a significantly larger region at higher masses given by 
\beq
   {m\over m_\phi} ~>~ \sqrt{ {1\over 4} - {1\over 2+\Delta \gamma} }~.
\label{highregime}
\eeq
For other ranges of $m$, the corresponding abundances decrease as a function of $m$.

This non-monotonic behavior gives great flexibility to the DDM model-builder:  
one need only select a model with ensemble constituent masses $m_i$ at appropriate locations 
along these curves
in order to endow 
these constituents with cosmological abundances 
which either rise or fall with mass, at will.
For example, if one wishes to have ensemble constituents whose cosmological abundances all fall monotonically with mass,
one need only choose these constituents to have masses $m_i$ within the ranges specified above
(or choose $\Delta \gamma \leq 2$, for which the corresponding ranges are unrestrained).   
Thus even the cases with $\Delta \gamma >2$ are capable of yielding purely monotonically falling abundances 
$\Omega_i$, as typically desired for DDM.~
However, the presence of non-monotonicities in these cases also allows for other possibilities.
For example, through appropriate choices of constituent masses $m_i$, one can imagine situations 
in which thermal freeze-out yields growing abundances for one dark-matter species
(or one portion of a DDM ensemble)
and yet decreasing abundances for another.
Indeed, by adjusting the values of the constituent masses $m_i$, one can even dial
the relevant value
of the scaling exponent $\gamma$ in a continuous way.
These observations
thus significantly enrich the phenomenological possibilities for 
DDM model-building.

Thus far, we have considered the cases in which our final-state particles $\psi$ are spin-0 or spin-1/2. 
However, it is also useful to consider the case in which $\psi$ is a spin-$1$ particle, 
henceforth to be denoted $A^\mu$.  For concreteness, let us 
imagine that $A^\mu$ is the gauge boson of a (potentially broken) gauge symmetry,
either abelian or non-abelian,
and let us endow $A^\mu$ with an arbitrary mass $m_\psi= m_A$.
For concreteness,
let us further take $\chi$ to be spin-1/2, our mediator $\phi$ to be spin-$0$, and our
$\phi A_\mu A^\mu$ coupling
to be of the form that emerges from the gauge-invariant operator
\beq
                { c_A\over \Lambda} \, \phi\, F_{\mu\nu}^a F^{\mu\nu a}~.
\label{nonren}
\eeq
Note that we adopt this gauge-invariant coupling structure in order to accommodate the
case with $m_A=0$, for which our gauge symmetry is necessarily unbroken.
Calculations similar to those above
then lead to a cosmological abundance $\Omega(m)$ whose scaling behavior
takes the form given in Eq.~(\ref{finalresult2})
with $s=t=0$, with $g_\psi\to c_A$ and $\epsilon_\psi \to \epsilon_A= -1$, and with the final-state kinematic factor
$(1-m_\psi^2/m^2)^{-1/2}$ now multiplied by an additional kinematic factor 
$f(x) \equiv (1- x^2 + 3x^4/8)^{-1}$ where $x\equiv m_A/m$.
For $x\ll 1$, we can approximate $f(x)\approx (1-x^2)^{-1}$,
which is tantamount to our traditional form in Eq.~(\ref{finalresult2}) with $s=1$ rather than $s=0$.  
But regardless of the value of $x$, we see that the primary effect of taking our final-state
particles to be spin-$1$ with the coupling indicated in Eq.~(\ref{nonren})
is that $\epsilon_\psi$ (now denoted $\epsilon_A$) is {\it negative}\/.  This is a direct consequence
of the fact that the leading-order gauge-invariant coupling between a spin-0 mediator and two vector fields,
as in Eq.~(\ref{nonren}), is non-renormalizable.
This then has the net effect of allowing situations with $\Delta\gamma = -2$,
which only further strengthens the desired inverse scaling between the abundances and masses
and which produces values for $\gamma$ 
which are even more negative than those which 
emerge for any other cases considered thus far.

If we restrict our attention to cases in which the vectors $A^\mu$ are necessarily massive,
then the gauge symmetry is necessarily broken and  
a fully gauge-invariant coupling such as that in Eq.~(\ref{nonren})
is not required.
In such cases, we may instead consider a direct super-renormalizable coupling of the form
\beq
                { c_A \mu} \, \phi\, A_\mu^a A^{\mu a}~,
\label{nonren2}
\eeq
which is reminiscent of the couplings for massive vector mediators in Tables~\ref{Tab:X} and \ref{Tab:psi}.
We then find that the corresponding cosmological abundance $\Omega(m)$ varies with $m$ 
exactly as it does for the $\phi FF$ coupling discussed above, except with
$f(x)$ now given by $f(x) \equiv (1- x^2 + 3x^4/4)^{-1}$.
Thus, for all intents and purposes, the change of coupling structure from that in Eq.~(\ref{nonren}) to
that in Eq.~(\ref{nonren2}) has very little effect on the resulting scaling of $\Omega$ with $m$. 
At first glance, it may seem surprising that we continue to have $\epsilon_A= -1$ when we are now
dealing with the super-renormalizable operator in Eq.~(\ref{nonren2}).
However, we can always algebraically recast our result into a form with $\epsilon_A= +1$ 
by replacing $f(x)$ with $g(x)\equiv  (3 x^4/4) f(x) = (1- 4 x^{-2}/3 + 4 x^{-4}/3)^{-1}$.
Of course, this algebraic manipulation does not change the underlying 
scaling behavior, which continues to be the same as that for the $\phi FF$ coupling.
We see, then, that the abundance function $\Omega(m)$ resulting from the coupling in Eq.~(\ref{nonren2}) has 
an almost identical scaling behavior 
as that resulting from the coupling in Eq.~(\ref{nonren}).
Indeed, in both cases 
the resulting scaling exponents $\gamma$ are more negative than for any other cases we have considered.

\FloatBarrier

%%%%%%%%%%%%%%%%%%%%%%%%%%%%%%%%%%%%%%%%%%%%%%%%%%%%%%%%%%%%%%%%%%%%%%%%%%%%%%%%%%%%%%

\section{Balancing lifetimes against abundances:  ~General constraints for DDM viability
\label{sec:DecayRateScaling}}

%%%%%%%%%%%%%%%%%%%%%%%%%%%%%%%%%%%%%%%%%%%%%%%%%%%%%%%%%%%%%%%%%%%%%%%%%%%%%%%%%%%%%%

In Sects.~\ref{sec:ThermalFreezeout} and \ref{general},
we have discussed the means by which we can achieve an ensemble of states
for which the cosmological abundances 
produced through thermal freeze-out
scale inversely with mass.
As discussed in the 
Introduction, and as we shall further discuss below,
this scaling behavior is a primary ingredient leading to a viable DDM ensemble.

However, this alone is not sufficient.  The DDM framework 
also requires certain scaling behaviors for the {\it decay widths}\/ $\Gamma_i$ of our ensemble
constituents, where we assume that the dominant decay mode of each DDM ensemble constituent 
is directly
into SM states.
Likewise, the DDM framework
also requires certain scaling relations for the {\it mass distribution}\/ of states across the ensemble,
or equivalently for the corresponding ensemble {\it density of states}\/.
Indeed, what ultimately matters for the phenomenological viability of a DDM ensemble
is how these different scaling behaviors balance against each other~\cite{DDM1}.
In this Section we shall briefly review the scaling relations for the decay widths and
densities of states.
We shall also outline some of the general constraints that they must satisfy, 
and what our results for the scaling behaviors of the 
cosmological abundances imply about these other scaling relations. 
 
In general, a given ensemble of dark-matter states
will typically have SM decay widths $\Gamma_i$ exhibiting simple scaling 
behaviors as functions of the constituent masses $m_i$.
For example, 
let us consider what is perhaps the simplest decay pattern in which each ensemble constituent $\chi_i$ decays
directly into two final-state particles $f$ and $\overline{f}$ whose 
masses are well below those of the constituents:  $m_f\ll m_i$ for all $i$.
We also assume that this decay occurs through a dimension-$d$ contact operator which therefore takes 
the form ${\cal O}_i \sim c \chi_i {\overline{f}}f/\Lambda^{d-4}$ where $\Lambda$ is an appropriate mass scale.
Since the matrix element ${\cal M}$ for a $1\to 2$ decay of this form must have mass dimension $+1$,
we then find through elementary dimensional analysis that the matrix element must scale with the constituent mass
as ${\cal M} \sim m_i^{d-3}$.
Since the decay width under such circumstances generally scales as $\Gamma_i\sim |{\cal M}|^2/m_i$,  
we then find that
\beq
       \Gamma_i ~\sim~ m_i^{2d-7}~.
\eeq
Thus the decay widths scale as a positive power of the mass for $d\geq 4$, but with a negative power for $d\leq  3$.
In general, we can write our decay-width scaling 
relation in the form
\beq
       \Gamma_i ~\sim~ m_i^y
\eeq
where $y$ is an appropriate scaling exponent.   Of course, for decays of the simple form described above,  
we have $y= 2d-7$.

It is also natural to consider ensembles of states in which the distribution of constituent masses $m_i$, 
and thus the corresponding effective 
densities of states $n_m$ per unit mass, also obey power-law scaling relations.
For example, in many concrete realizations of 
DDM ensembles one finds that the constituent masses are distributed as
\beq
            m_k ~\sim~ k^\delta
\label{massrelation}
\eeq
where $\delta>0$ is another scaling exponent.
Special benchmark cases include $\delta=1$ (corresponding to the KK spectrum $m_k\sim k/R$ resulting from compactification on a circle or orbifold of radius $R$~\cite{DDM1,DDM2,DDMAxion}) 
as well as $\delta=1/2$ (corresponding to the spectrum $\alpha' m^2_k \sim k$ of string excitations~\cite{DDMHadrons,anupam},   where $\alpha'$ is the Regge slope, \ie,  the inverse of the squared string scale, and where $k$ is the string excitation number).  A general scaling relation of the form in Eq.~(\ref{massrelation}) for the ensemble mass spectrum 
then implies a corresponding density of states $n_m$ per unit mass which scales with mass as
\beq
                n_m ~\sim~ m^{1/\delta -1}~.
\eeq   

We thus have three independent scaling relations that govern the structure of our DDM ensemble:  
$\Omega\sim m^\gamma$, $\Gamma\sim m^y$, and $m_k\sim k^\delta$.
Corresponding to these are three scaling coefficients:  $\gamma$, $y$, and $\delta$.
In general, there are many detailed phenomenological constraints that govern the allowed values of these exponents.
However, for our purposes in this paper we shall concentrate on only the two most fundamental 
constraints that
ensure the ``zeroth-order'' phenomenological self-consistency of the DDM ensemble as a whole.
Indeed, our purpose in this Section is not to develop a detailed phenomenological
set of bounds on these scaling exponents so much as to understand the general
architecture of how these constraints play against each other across
the DDM ensemble.
A more detailed study of the phenomenological constraints on these scaling relations can be found in Ref.~\cite{pat}.

Our first constraint concerns the decay widths, or equivalently the lifetimes, of our ensemble states. 
In general,
any dark-matter particle which decays too rapidly into SM
states is likely to upset Big-Bang nucleosynthesis (BBN) and 
light-element abundances,
and also leave undesirable imprints in the cosmic microwave background (CMB) and
diffuse photon backgrounds.
However, if such a decaying particle carries a sufficiently small cosmological 
abundance at the time of its decay,
the disruptive effects of this decay will be minimal and all
constraints from BBN, the CMB, {\it etc.}\/, can potentially be satisfied.
This then leads to the fundamental notion~\cite{DDM1} which serves as the underpinning of the DDM framework,
namely that dark-matter {\it stability}\/ is no longer required in a multi-component context
so long as it is replaced by a {\it balancing}\/ of lifetimes against abundances across
the dark-matter ensemble, 
so that states carrying larger abundances are longer-lived while states
that are shorter-lived carry smaller abundances.  
This in turn requires that abundances scale inversely with decay widths, \ie, that
\beq
    \Omega_i~\sim~ \Gamma_i^\alpha~~~~~{\rm  where}~~ \alpha<0~. 
\label{alphadef}
\eeq
With $\Omega\sim m^\gamma$ and $\Gamma\sim m^y$ we find that $\alpha= \gamma/y$, whereupon
Eq.~(\ref{alphadef}) yields
\beq
            \gamma/y ~< ~ 0~.
\label{gammayconstraint}
\eeq
Thus $\gamma<0$ corresponds to  $y>0$, which for
a two-body decay into light fermions leads to the  constraint $d\geq 4$,
as discussed above.

Strictly speaking, 
the constraint in Eq.~(\ref{gammayconstraint}) should be understood as applying
to only those portions of the DDM ensemble consisting of dark-matter 
components whose decays into SM states
have the potential to be phenomenologically problematic.
For example, extremely heavy dark-matter states decaying during extremely early periods of cosmological evolution
well before BBN need not satisfy these bounds, as 
the decay products rapidly thermalize with the radiation bath.  
Consequently, while such decays can potentially
induce a later period of reheating,
depending on the abundances of the decaying particles~\cite{Watson}, 
they typically have few other 
observable consequences.
Thus one could conceivably tolerate having $\gamma,y>0$ 
within such portions of the ensemble.  
This issue will be discussed in detail in Ref.~\cite{pat}.
Likewise, the cosmological abundances of dark-matter states whose  
lifetimes significantly exceed $10^9 t_{\rm now}$ are also unconstrained and also may exhibit $\gamma,y >0$.
Thus, it is only within that all-important region of the DDM ensemble consisting of states
with lifetimes $\tau$ in the range $t_{\rm BBN} \lsim \tau \lsim 10^9 t_{\rm now}$ that
we must demand $\gamma/y<0$.

It is also important to note that in deriving the result in Eq.~(\ref{gammayconstraint}), we have
not assumed that $\gamma$ is a constant throughout the relevant portion of the DDM ensemble.
Likewise, we have also not assumed that $y$ is a constant over this range.
Instead, 
we simply need to verify that Eq.~(\ref{gammayconstraint}) holds
throughout the relevant portion of the ensemble.
Indeed, both $\gamma$ and $y$ are free to vary so long as
the constraint in Eq.~(\ref{gammayconstraint}) is satisfied.

Our second constraint on these scaling exponents concerns the time-development of the total dark-matter
abundance 
$\Omega_{\rm tot} \equiv \sum_i \Omega_i$, 
or equivalently the time-development of the corresponding energy density 
$\rho_{\rm tot}$,
where 
$\Omega_{\rm tot} = \rho_{\rm tot}/\rho_{\rm crit}$.
Here $\rho_{\rm crit}= 2 M_P^2 H^2$, where
$H$ is the Hubble parameter and where $M_P\equiv (8\pi G_N)^{-1/2}$
is the reduced Planck scale. If each constituent within
the ensemble were stable, the total energy within the
ensemble would remain constant, implying that the energy 
density $\rho_{\rm tot}$ would fall as a function of time solely
because of the Hubble expansion of the universe, with
$d\rho_{\rm tot}/dt = -3H\rho_{\rm tot}$.
In other words, the collective
equation-of-state parameter $w_{\rm eff}$ for the dark-matter 
ensemble as a whole, defined as~\cite{DDM1}
\beq
       w_{\rm eff}(t) ~\equiv~ -\left( {1\over 3H\rho_{\rm tot}}  {d\rho_{\rm tot}\over dt} + 1\right) ~,
\label{weffdef}
\eeq
would vanish. The vanishing of $w_{\rm eff}(t)$ under such stability assumptions is of course consistent with the 
interpretation of the corresponding energy density of our
ensemble as being associated with dark {\it matter}\/ (as opposed to dark energy or dark radiation), which is in turn
consistent with observational constraints. However, the
constituents within our dark ensemble are not stable: as
described above, they decay with lifetimes that obey certain scaling relations relative to 
their cosmological abundances.  These decays in turn cause 
$\Omega_{\rm tot}$ to fall as a function of time, 
leading to a positive value of $w_{\rm eff}$.  The
scaling relations that govern these decays must therefore
be balanced in such a way~\cite{DDM1}  that $w_{\rm eff}$  not
be too far from
zero at the present time and also not have varied significantly 
within the recent cosmological past. 
The first of these requirements ensures that we can continue to interpret the energy within the 
DDM ensemble as associated with dark matter, within experimental constraints.
By contrast, 
the second requirement stems from the observation that 
the behavior of $w_{\rm eff}(t)$ has an impact on the expansion history of the universe,  
independent of the constraint in Eq.~(\ref{gammayconstraint}). 
Indeed, the behavior of $w_{\rm eff}(t)$ is constrained~\cite{BlackadderKoushiappas1,BlackadderKoushiappas2} by a combination of CMB data~\cite{WMAP,Planck}; observations of baryon acoustic oscillations in galaxies~\cite{GalacticBAOBOSS,GalacticBAO6dF,GalacticBAOSDSS} and in the Lyman-$\alpha$ forest~\cite{LymanAlphaBAOBOSS1,LymanAlphaBAOBOSS2}; and measurements of the redshifts and luminosity distances of Type-Ia supernovae~\cite{SupernovaTypeIaSDSS}.
Moreover, modifications to the expansion rate of the universe can also affect the light-element abundances
generated during the BBN epoch.
Thus, for all practical purposes it is reasonable to identify the ``recent past'' over which 
$w_{\rm eff}(t)$ should not vary significantly as the period since BBN.

In this connection, we note that an equation of state is
a property intrinsic to the ensemble. As such, $w_{\rm eff}(t)$
is independent of the background cosmological epoch.
Thus the constraints regarding the behavior of $w_{\rm eff}(t)$
that we have indicated above are applicable regardless
of whether we are considering a matter- or radiation-dominated epoch, 
and likewise also apply across transitions between epochs.

These constraints were investigated in Ref.~\cite{DDM1}.
It turns out that 
the constraint that $w_\ast\equiv w_{\rm eff}(t_{\rm now})$ not be too far from zero
is ultimately independent of the scaling exponents and only requires 
suitable
overall {\it normalizations}\/ for our abundance and lifetime scaling relations~\cite{DDM1}. 
In other words, the value of $w_\ast$ ultimately depends on the overall prefactor coefficients
that come into these scaling relations but not on the scaling exponents themselves.
We shall therefore henceforth assume that this constraint has been satisfied
and that $w_\ast$, though positive, is extremely small and ultimately within the experimental
bounds consistent
with an interpretation in terms of dark matter.

By contrast, requiring that $w_{\rm eff}(t)$ not have varied 
significantly within the recent past
leads directly to a constraint on our scaling exponents~\cite{DDM1}.
Following the discussion in Ref.~\cite{DDM1}, this constraint may be phrased
as follows.
In the limit that our ensemble consists of a large number of densely-packed dark-matter states,
we can imagine that the
spectrum of discrete decay widths $\Gamma_i$ 
is nearly continuous, parametrized by a continuous variable $\Gamma$.  
In this approximation, we can view the spectrum of abundances $\Omega_i$ as a continuous
function $\Omega(\Gamma)$ of decay widths.
We can likewise express our density of states as a density of states per unit $\Gamma$, henceforth
denoted $n_\Gamma$.
In general, both of these quantities will have scaling behaviors of the form
\beq
        \Omega(\Gamma)~\sim~ \Gamma^\alpha~,~~~~~~~ n_\Gamma~\sim~\Gamma^\beta~,
\eeq
where $\alpha$ is the same exponent we have already seen in Eq.~(\ref{alphadef}).
It then turns out~\cite{DDM1} that the corresponding 
equation-of-state parameter $w_{\rm eff}(t)$ depends only 
on $w_\ast$ and on the sum $x\equiv \alpha+\beta$.  Indeed, $w_{\rm eff}$ is generally a rather non-trivial 
function of these variables.
However, for $w_\ast\ll 1$, one finds~\cite{DDM1}
\beq
      w_{\rm eff}(t) ~\approx~ w_\ast \left( {t\over t_{\rm now}}\right)^{-x-1}~.
\label{wefft}
\eeq
Given that $w_\ast\ll 1$, we thus see that we must have $x\leq -1$ in order for $w_{\rm eff}(t)$ to have remained
small throughout the recent past. 
Again we stress that this conclusion holds even though our definition of ``recent past''
stretches across both radiation- and matter-dominated epochs.

Of course, strictly speaking, any value of $x\leq -1$ is permitted for those ensemble states decaying 
within this time interval, even values of $x$ which are significantly less than $-1$. 
Taking $x\ll -1$ simply means that even
though $w_{\rm eff}(t)$ has a very small value $w_\ast\ll1$ at the present epoch,
it approaches zero extremely rapidly as we go backwards in time towards BBN.~  
However, while a choice $x\ll -1$ succeeds in guaranteeing 
$w_{\rm eff}(t)\ll 1$ during the entire recent period since BBN, such a choice is unnatural from several points
of view.
First, if $x\ll -1$ for all ensemble states decaying between BBN and the present epoch, 
it is natural to assume that similar values of $x$ would continue
to hold for states immediately beyond this range, \ie, for states whose 
decays will occur in the immediate future.
However, this would then cause  
$w_{\rm eff}(t)$ to experience a sudden dramatic growth  for $t>t_{\rm now}$. 
While this cannot be ruled out solely on the grounds of equations of state and their behaviors, 
such a sudden dramatic change in $w_{\rm eff}(t)$ beyond $w_{\rm eff}\ll 1$
would effectively single out the present time as a special epoch in the cosmological timeline.
Indeed, the only way to avoid this would be to assume that $x$ itself must experience a sudden change
at the beginning of that portion of the ensemble whose states decay in the present epoch.
However, this too would single out the present epoch as special. 
There may also be more direct phenomenological reasons to exclude having $x\ll -1$.
Since non-zero values of $w_{\rm eff}(t)$ are ultimately due to the decays of our dark-matter constituents,
a sudden, rapid growth in $w_{\rm eff}(t)$   
is likely to be correlated with a large injection of decay products, potentially including large amounts of radiation.
If these decay products include SM particles, this in turn is 
likely to cause issues with diffuse photon backgrounds, {\it etc.} 
This too will be discussed in more detail in Ref.~\cite{pat}.

Thus, for such reasons, it is more natural 
to assume that \hbox{$x{\lsim} -1$}.   In other words, we shall henceforth assume that $x$, though less than $-1$,
is not too far below $-1$.  We shall demand that this be true over that portion of the ensemble
whose states decay between BBN and  the present epoch.  This assumption allows us to avoid sudden changes in either
 the cosmological evolution or the structure of the ensemble once the current epoch is reached.

It is straightforward to express the constraint $x\lsim -1$ in terms of our 
scaling coefficients $\lbrace \gamma, y, \delta\rbrace$.
We recall that $x\equiv \alpha  + \beta$, and we have already seen below 
Eq.~(\ref{alphadef})
that $\alpha= \gamma/y$.
To calculate $\beta$, we observe that
\beq
  n_\Gamma = n_m \left|\frac{dm}{d\Gamma}\right|\sim m^{1/\delta -1} m^{1-y}
       \sim \Gamma^{(1/\delta-1)/y} \Gamma^{(1-y)/y}~,  
\eeq
allowing us to identify $\beta = 1/(y\delta) -1$.
We thus find that
\beq
          x ~=~ {\gamma\over y} + {1\over y\delta} -1~,
\eeq
whereupon the constraint $x\lsim -1$ reduces to
\beq
         {1\over y} \left( \gamma + {1\over \delta}\right) ~\lsim~0~.
\eeq
For positive $y$ we thus find 
\beq
              y>0:~~~~~~\gamma ~\lsim~ -1/\delta~,
\label{sinceBBN}
\eeq
while for negative $y$ we find
\beq
              y<0:~~~~~~\gamma ~\gsim~ -1/\delta~.
\eeq
In either case, these inequalities can be rewritten in the common form
\beq
        \delta ~\gsim~ \delta_{\rm min}\equiv -1/\gamma~.
\label{sinceBBN2}
\eeq
Indeed, this constraint holds regardless of the sign of $y$.

Remarkably, we see that the scaling exponent $y$ has completely dropped out of this constraint.
Moreover, we observe that this constraint algebraically has the same form as the constraint that
would emerge from demanding that $\Omega_{\rm tot}$ be finite in cases where our ensemble consists
of an infinite tower of states whose masses stretch to infinity.  In such cases we would have
\beq
      \Omega_{\rm tot} ~=~  \int dm\, n_m\, \Omega(m) ~\sim~  \int dm \, m^{1/\delta-1} m^{\gamma}~,~~~~
\label{integralsfinite1}
\eeq
and we see that the ``ultraviolet'' finiteness of this integral requires that $\gamma \leq -1/\delta$.
Of course, despite their algebraic similarity, at a physical level these are ultimately different constraints since
the constraint stemming from Eq.~(\ref{integralsfinite1})   
applies for either sign of $y$  and
applies only for those 
states at the large-mass ``ultraviolet'' end of the ensemble, 
while the constraint in Eq.~(\ref{sinceBBN}) 
assumes that $y$ is positive and 
needs only apply within that portion of the ensemble whose
states decay between BBN and the present epoch.
It is nevertheless interesting that both constraints, operating over different portions of the ensemble,
share a common algebraic structure.

Just as with the constraint in Eq.~(\ref{gammayconstraint}),
we emphasize 
that Eq.~(\ref{sinceBBN2})
must hold only
throughout the relevant portions of the ensemble discussed above.
Indeed, both $\gamma$ and $\delta$ are free to vary 
so long as Eq.~(\ref{sinceBBN2}) holds at each mass scale within this region. 

As indicated above, the constraint in Eq.~(\ref{sinceBBN2}) can 
be satisfied with $\gamma$ either positive or negative.
Indeed, in either case we need only require 
that $y$ and $\gamma$ have opposite signs, 
in accordance with Eq.~(\ref{gammayconstraint}).
However, these results help to explain why the situation with $\gamma<0$ is more natural from a 
DDM perspective.
Note that $\delta$ is necessarily positive, since our states are
ordered in terms of increasing mass by construction. 
Moreover, when $\gamma$ is negative, we find that $\delta_{\rm min}$ is positive.
This implies that $\delta$ can indeed easily satisfy $\delta\gsim\delta_{\rm min}$ --- {\it i.e.}\/,
it is not difficult for $\delta$ to be only slightly greater than $\delta_{\rm min}$.   
For $\gamma<0$, by contrast, $\delta_{\rm min}$ is negative.
Thus,  although {\it any}\/ positive value of $\delta$ is greater than $\delta_{\rm min}$, 
it can be difficult for $\delta$ to be both positive and only slightly greater than $\delta_{\rm min}$.
Indeed, this latter requirement becomes increasingly hard to satisfy when $\gamma$ is small.
There are also other reasons to prefer $\gamma<0$.
In general, it is more natural for heavier states to have smaller lifetimes
and larger decay widths than lighter states.
This requires $y>0$, which in turn requires $\gamma<0$.
Indeed, it is for all of these reasons that it has been
crucial to find ways in which we might obtain abundances
with $\gamma < 0$ from thermal freeze-out. Of course, within
other regions of the ensemble, no such constraints need apply.
Such regions could then have 
$\gamma > 0$, even while $\delta,y > 0$.

%---------------------BEGIN FIGURE-----------------------%
\begin{figure}[h!]
\includegraphics[keepaspectratio, width=0.42\textwidth]{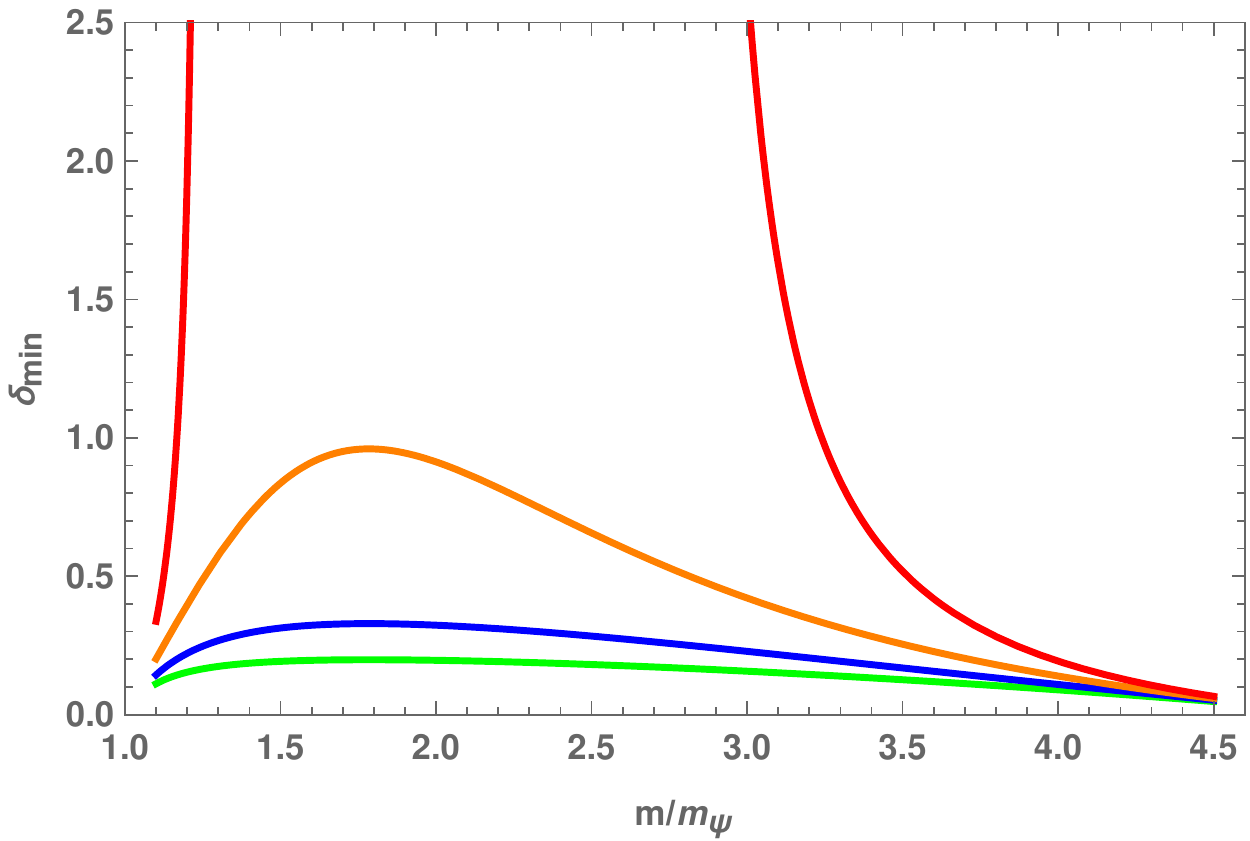}
\includegraphics[keepaspectratio, width=0.42\textwidth]{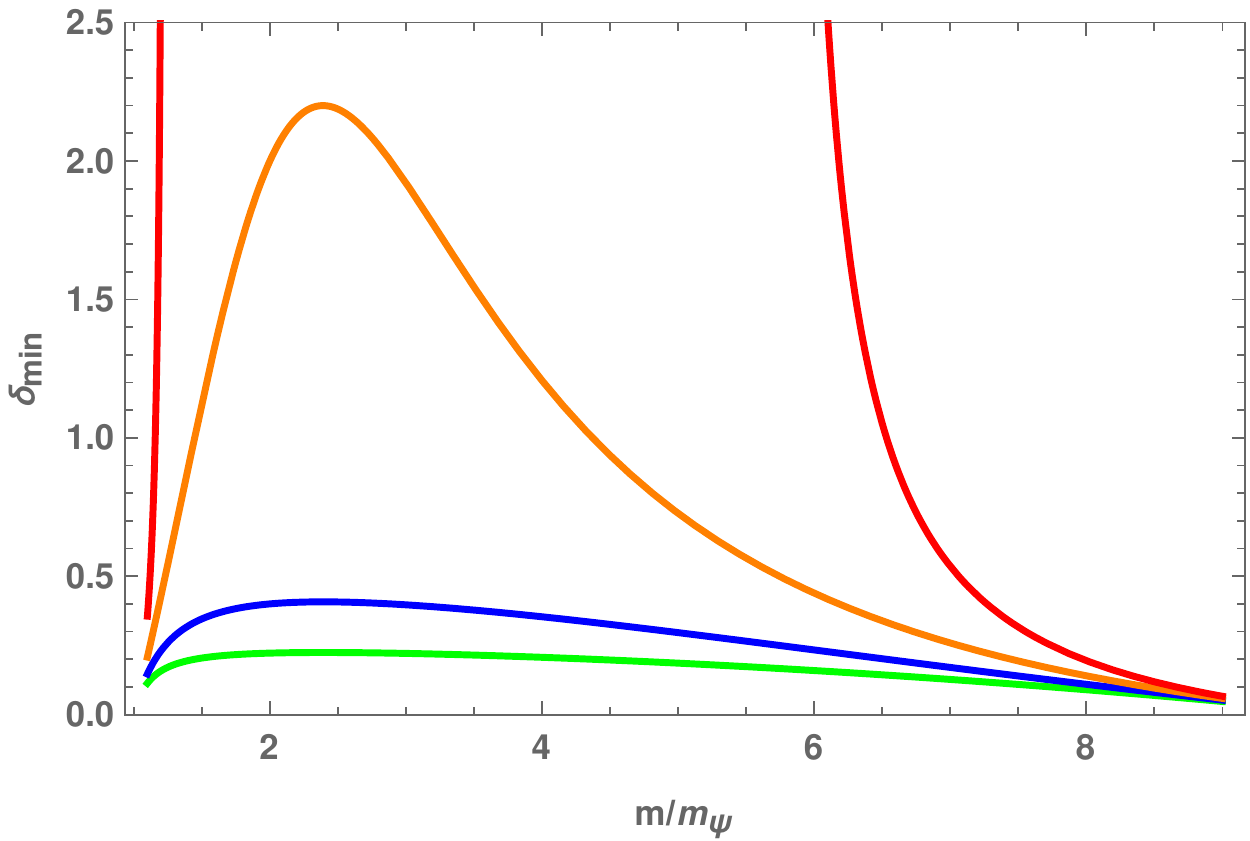}
\includegraphics[keepaspectratio, width=0.42\textwidth]{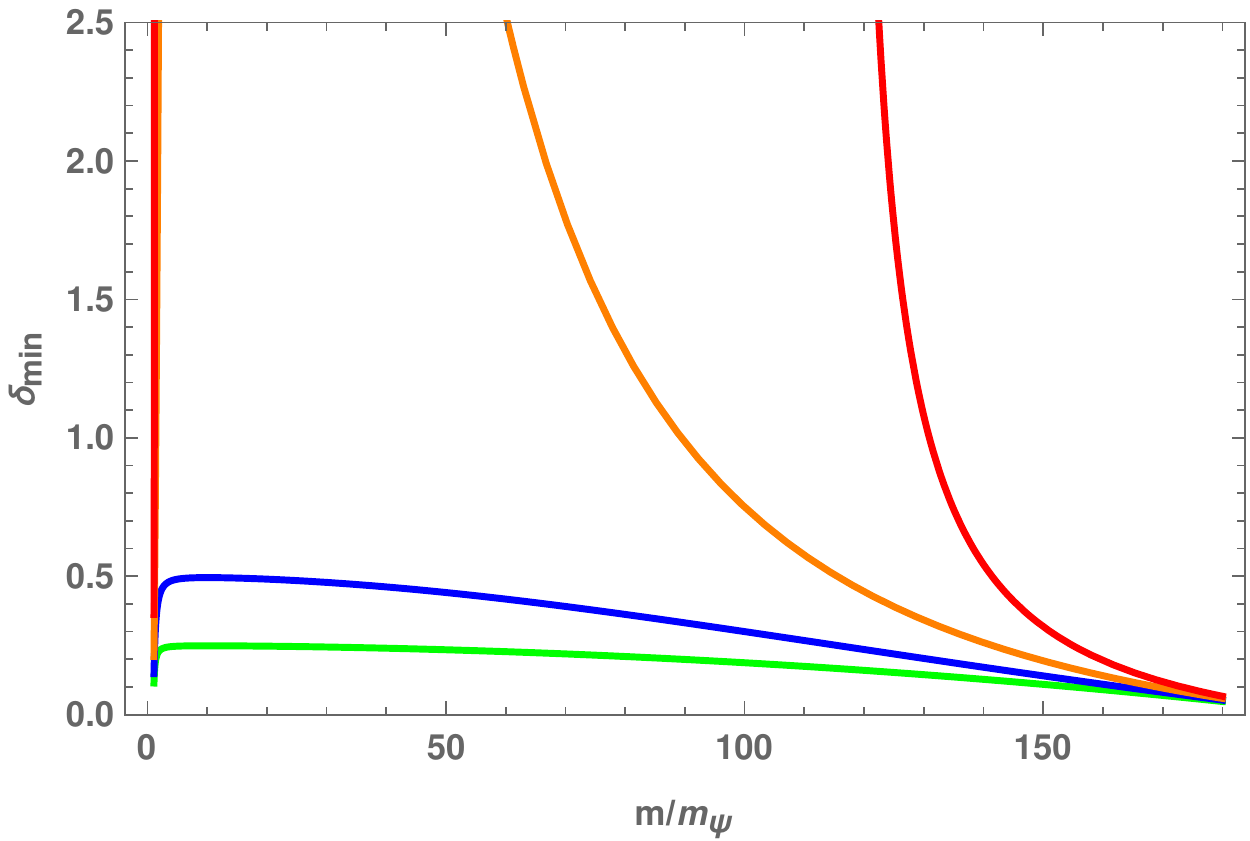}
\caption{Minimum $\delta$-exponents $\delta_{\rm min}\equiv -1/\gamma$, plotted as functions
of $m/m_\psi$ for $m_\phi/m_\psi=10$ (top panel), $20$ (middle panel), and $400$ (bottom panel).
Each panel shows results for $\Delta\gamma = -2$ (green), $0$ (blue), $2$ (orange), and $4$ (red);
for $\Delta\gamma\geq 0$ these results correspond to the $\Omega$ plots shown in 
Figs.~\ref{othercases} and \ref{othercases2}.
For $\Delta \gamma < 2$, we see that $\delta_{\rm min}$ remains finite for all $m_\phi/m_\psi$.
This remains true even in
the $\Delta\gamma=2$ case,
for which there exist certain mass regions 
in which $\delta_{\rm min}$ 
grows significantly as $m_\phi/m_\psi\to\infty$, reflecting the increasing tendency of the corresponding
abundances $\Omega$ to become almost flat as functions of $m$.
For $\Delta \gamma =4$, by contrast, $\delta_{\rm min}$ actually diverges at 
the two masses for which the corresponding
values of $\Omega$ in Figs.~\ref{othercases} and \ref{othercases2} reach local minima or maxima. 
Between these masses $\gamma>0$ and the constraint in Eq.~(\ref{sinceBBN2}) becomes vacuous.}
\label{fig:deltamin}
\end{figure}
%----------------------END FIGURE------------------------%

Turning back to Eq.~(\ref{sinceBBN2}),
we see that
for each value of $\gamma$ 
there is a corresponding minimum mass-distribution scaling exponent
$\delta_{\rm min}$ that is permissible in order to ensure
that $x\lsim -1$.
In Fig.~\ref{fig:deltamin}, we have plotted $\delta_{\rm min}$ as a function of $m$
for different values of $m_\phi/m_\psi$.
In each case we have plotted results for $\Delta\gamma= 0,2,4$, corresponding to the
curves in Figs.~\ref{othercases} and \ref{othercases2}, and we have also shown
results for $\Delta\gamma = -2$, corresponding to the case
in which thermal freeze-out occurs due to dark-matter annihilation
into pairs of spin-1 particles.
We immediately see that for the entire allowed mass range,
the cases with $\Delta \gamma= -2$ and $\Delta\gamma =0$ are consistent
with not only $\delta= 1$, corresponding to a DDM ensemble consisting of the KK states
resulting from circle compactification,
but also $\delta= 1/2$, corresponding to a DDM ensemble of string-like
origin.
However we also learn that these constraints become more exclusive
as $\Delta \gamma$ increases, and tend to allow such values of  
$\delta$ only within certain mass regions.   For example, we see that an ensemble
of KK states is consistent with the abundances produced through a thermal production
process with $\Delta\gamma = 2$ and $m_\phi/m_\psi \gg 1$ only within the approximate
mass range $m/m_\psi \gsim 100$.
Finally, we note that $\delta_{\rm min}$ is negative within those mass ranges
for which $\gamma >0$ in Figs.~\ref{othercases} and \ref{othercases2}.
The constraint 
in Eq.~(\ref{sinceBBN2}) is then vacuous and no value of $\delta_{\rm min}$ is plotted
in Fig.~\ref{fig:deltamin}.

Depending on which portions of the allowed mass ranges are actually populated
by the constituents of our DDM ensemble, 
we see that entirely different dark-matter phenomenologies can emerge.
As a result of the constraints in Eqs.~(\ref{gammayconstraint}) and (\ref{sinceBBN2}),
we see that the different classes of phenomenologically allowed thermal freeze-out processes,
as well as different classes of dark-matter decay processes,
are closely tied to the structure of (and indeed the 
corresponding distribution of masses across) the DDM ensemble.
This then provides important correlations between the particle physics of the ensemble structure,
the particle physics of the annihilations of its constituents,
and the overall (in this case, thermal) cosmological history in which this
particle physics is embedded.

\FloatBarrier

%%%%%%%%%%%%%%%%%%%%%%%%%%%%%%%%%%%%%%%%%%%%%%%%%%%%%%%%%%%%%%%%%%%%%%%%%%%%%%%%%%%%%%

\section{Discussion and  Conclusions \label{sec:Conclusions}}

%%%%%%%%%%%%%%%%%%%%%%%%%%%%%%%%%%%%%%%%%%%%%%%%%%%%%%%%%%%%%%%%%%%%%%%%%%%%%%%%%%%%%%

Within the DDM framework, the phenomenological viability of the dark-matter ensemble 
is the result of the interplay between three fundamental relations which govern 
how the masses, cosmological abundances, and decay widths of the individual ensemble
constituents scale relative to one another across the
ensemble as a whole.  
While the scaling relations for masses and decay widths
primarily depend on particle-physics considerations alone, the scaling relation for
cosmological abundances typically depends on a mix of particle-physics
and cosmological considerations.  Thus, any concrete realization of 
the DDM framework relies on there being an appropriate abundance-generation mechanism 
which not only arises naturally within the corresponding cosmological model but also gives rise 
to an abundance function $\Omega(m)$ with a suitable scaling behavior.

Most prior work realizing explicit DDM ensembles has   
relied on non-thermal abundance-production mechanisms such as vacuum misalignment, as these can easily give
rise to scaling relations in which the abundances scale inversely with mass.
As we have discussed, phenomenological constraints tend to prefer 
this behavior over large portions of any DDM ensemble.
Unfortunately, the simplest thermal freeze-out mechanisms tend to 
result in abundances with the opposite behavior, growing as a function of mass.
Indeed, this behavior is an intrinsic element underpinning the so-called ``WIMP miracle''.
The purpose of this paper has therefore been to determine whether the scaling behavior
desired for DDM ensembles might also be realized through thermal freeze-out.
This would in turn determine whether the DDM framework can be extended
into the thermal domain.

The results of this paper provide an answer in the affirmative.
Indeed, we have shown 
that relatively straightforward modifications to standard dark-matter annihilation
processes
result in freeze-out abundances which fall, rather than rise, 
as functions of constituent mass.
In fact, we have found that a whole spectrum of behaviors is possible,
with abundances that can fall within certain portions of our DDM ensemble and rise within others.
Indeed, by adjusting the values of the discrete dark-sector mass spectrum $\lbrace m_i \rbrace$, 
we can even continuously dial our abundance scaling exponents $\gamma$ across a wide
range of values. 
The results of this paper thus suggest that a rich and flexible dark-matter phenomenology
can result when the DDM framework is extended into the thermal domain.
This in turn can provide a versatile 
tool for model-building within the DDM framework.  
Indeed, we note that the typical mass window for successful thermal 
freeze-out is $\mathcal{O}(1)\mbox{~keV} \lesssim m \lesssim 100\mbox{~TeV}$.  Since 
this is also the range of mass scales for which direct-detection experiments and 
collider searches for missing transverse energy are typically sensitive, 
the kinds of DDM ensembles whose constituents receive their abundances from thermal 
freeze-out are also the kinds of ensembles capable of giving rise to the wealth of 
distinctive phenomenological signatures discussed in 
Refs.~\cite{DDMLHC1,DDMLHC2,DDMDD} --- signatures which can potentially serve to 
distinguish these ensembles from traditional dark-matter candidates.
Thus we expect {\it thermal}\/ DDM ensembles 
to have immediate implications for these kinds of experimental signatures and bounds ---
even potentially more than for their non-thermal cousins.

In general, a collection of dark-matter states 
will sequentially undergo freeze-out in order of their masses $m_i$, with the heaviest states
freezing out first.
Indeed, freeze-out of a given particle generally occurs when the temperature $T$ of the universe
is approximately the mass of the particle, with $x_i\equiv m_i/T\approx {\cal O}(20)$.
However, there are subleading logarithmic effects which allow
certain states to freeze out with slightly smaller values of $x_i$ than others, and these
subleading
logarithmic effects depend on the cross-sections $\langle \sigma_i v\rangle$.
It is these effects which ultimately determine the cosmological abundances with which
the states emerge after freeze-out.  
Thus, while the masses $m_i$ of the states in a given ensemble
determine the order of freeze-out as a function of {\it time}\/,
it is the corresponding cross-sections $\langle \sigma_i v\rangle$ 
which determine the order of freeze-out as a function of $x$. 
These are, of course, general statements concerning the physics of the freeze-out process,
and they remain true in our scenarios as well.
However, what we have shown in this paper is that 
while we cannot adjust the order in which our ensemble constituents freeze out as a function of time,
we can certainly adjust the order in which they freeze out as a function of $x$.
Indeed, what we have shown is that
there exist annihilation processes for which the cross-sections $\langle \sigma_i v\rangle$
induce the {\it lighter}\/ states to freeze out with {\it smaller}\/  values of $x$, 
thereby imparting larger abundances to these states and 
producing the desired negative scaling exponent $\gamma<0$.
Moreover, as we have seen, it is even possible to arrange our states to freeze out 
in a {\it non-monotonic order}\/ as a function of $x$, 
with the states sitting at the local maxima of our abundance curves in Figs.~\ref{othercases}
and \ref{othercases2} freezing out at the smallest values of $x$ 
and those at local minima freezing out at the largest.

We are not the first to demonstrate that thermal freeze-out can yield abundance scaling 
relations that differ from those associated with the traditional WIMP paradigm.
Recall that standard thermal freeze-out during a 
radiation-dominated (RD) epoch yields a dark-matter abundance which scales with
the freeze-out temperature $T_f$, the mass $m_\chi$ of the dark-matter particle,
and the thermally-averaged annihilation cross-section $\langle\sigma v\rangle$
according to the relation
\begin{equation}
  \Omega_\chi ~\propto~ \frac{m_\chi}{T_f \langle\sigma v\rangle}~.
\end{equation}
The ratio $m_\chi/T_f$ is famously independent of $\langle\sigma v\rangle$ up to 
logarithmic corrections.  Thus, in traditional WIMP scenarios, in which 
$\langle\sigma v\rangle \propto g_\chi^4/m_\chi^2$, 
one recovers Eq.~(\ref{WIMPmiracle}).
However, in scenarios with non-standard cosmological histories in which
dark-matter freeze-out does not take place during an RD epoch, this scaling
relation is altered.  One example is the case in which freeze-out occurs immediately
prior to a late period of reheating --- \ie, during an epoch in which the universe
is dominated by a non-relativistic particle species (or by the zero mode of a 
rapidly oscillating scalar field) which is continually decaying into radiation.
In such scenarios, provided that $\langle\sigma v\rangle$ is sufficiently large that
thermal equilibrium is established between the dark-matter particles and the radiation
bath, one finds~\cite{FreezeoutInMDGiudice,FreezeoutInMDErickcek}
\begin{equation}
  \Omega_\chi ~\propto~ \frac{m_\chi T_{\mathrm{RH}}^3}{T_f^4\langle\sigma v\rangle}~,
\end{equation}
where $T_{\mathrm{RH}}$ is the reheat temperature associated with this late period
of reheating.  By contrast, if $\langle\sigma v\rangle$ is small and the dark matter
never thermalizes, the dark matter ``freezes in''~\cite{FreezeIn} rather than freezing
out and one finds~\cite{FreezeoutInMDErickcek}     
\begin{equation}
  \Omega_\chi ~\propto~ \frac{T_{\mathrm{RH}}^7\langle\sigma v\rangle}{m_\chi^5}~.
\end{equation}
Likewise, if thermal freeze-out occurs during an epoch in which the energy density of 
the universe is dominated by the kinetic energy associated with a rapidly rolling
scalar-field zero mode (as in so-called ``kination'' 
scenarios~\cite{SpokoinyKination,JoyceKination,FerreiraKination}), one finds~\cite{RedmondErickcekKination}
\begin{equation}
  \Omega_\chi ~\propto~ \frac{m_\chi}{T_{\mathrm{RH}} \langle\sigma v\rangle}~,
\end{equation}
up to logarithmic corrections in $T_f/T_{\mathrm{RH}}$.
All of these scenarios alter 
the standard $m_\chi$-dependence for the resulting cosmological abundance.

There are also a variety of {\it non}\/-thermal mechanisms through which a sizable dark-matter abundance 
can be generated. As one might expect, these mechanisms 
lead to altogether different scaling relations for $\Omega_\chi$ as a function of $m_\chi$.
One such mechanism is production through the decays of some other, heavier particle
which comes to dominate the energy density of the universe at early times.
In the regime in which the dark-matter particles produced in this way are effectively
decoupled from the radiation, the contribution to $\Omega_\chi$ is proportional~\cite{GelminiNonthermal} 
to the fraction $f_\chi$ of the energy of the heavy decaying particle that is transferred to 
$\chi$ (rather than to other decay products)
and is approximately independent of $m_\chi$.  By contrast, in the opposite regime in which
$\langle\sigma v\rangle$ is sufficiently large that the dark-matter particles produced 
by such decays undergo significant annihilation, the number density $n_\chi$ of
such particles is depleted by annihilation to 
$n_\chi \sim H(T_{\mathrm{RH}})/\langle\sigma v\rangle$, where $H(T_{\mathrm{RH}})$ is the 
value of the Hubble parameter at the reheat temperature associated with the late period
of reheating induced by the decay of the heavy particle.  In this case, one 
finds~\cite{ArcadiUllio}
\begin{equation}
  \Omega_\chi ~\propto~ \frac{m_\chi}{T_{\mathrm{RH}} \langle\sigma v\rangle}~.
\end{equation}
Of course,
other non-thermal mechanisms for abundance generation exist as well.  
These include, for example,
misalignment production as well as production via the decays of topological
defects (cosmic strings, domain walls, \etc) or non-topological 
solitons (\eg, oscillons~\cite{BogolyubskyOscillon,GleiserOscillon,EdOscillon}).  
These abundance-generation mechanisms each have their
own characteristic scaling relations with $m_\chi$ and with the other relevant 
parameters of the theory. 

These examples illustrate the ways in which modified cosmologies
can produce a variety of possible scaling 
relations between the cosmological abundance $\Omega_\chi$ 
and quantities such as $m_\chi$ and $\langle\sigma v\rangle$. 
Indeed, by invoking an early period of matter domination, 
kination, \etc, one can achieve almost any scaling behavior one desires.  However,
we have shown in this paper
that such departures from the standard cosmology are not necessary in order to obtain the appropriate 
scaling relations for a DDM ensemble through thermal freeze-out.  Indeed, we have
shown that the desired relations arise naturally within the standard cosmology and 
are realized in a particularly simple class of particle-physics models. 
In other words,
{\it our approach to modifying the traditional scaling relations expected from thermal freeze-out
has involved modifying the particle physics rather than modifying the cosmological narrative
in which the particle physics is embedded.}

Our results in this paper suggest many areas for further research.
First,  it would be interesting to explore the phenomenology that might result from thermal freeze-out
due to other, more complex annihilation processes.
For example, given that the DDM framework involves large multiplicities of dark-matter states,
it might be interesting to consider a strongly 
interacting massive particle (SIMP) framework~\cite{SIMP,SIMPWacker}
within which dark-matter annihilation might receive significant contributions from 
$3\to 2$ and perhaps even $4\to 2$ processes in which dark-sector particles 
annihilate into other dark-sector particles.
Likewise, in this paper we have assumed that our dominant annihilation mode 
is one in which our dark-sector ensemble components $\chi_i$ annihilate to states $\psi$ which
are outside the  ensemble.   However, in a multi-component framework such as the DDM framework,
there is also the possibility of {\it intra-ensemble}\/ 
annihilation processes of the form $\chibar_i\chi_i\to \chibar_j\chi_j$
where $\chi_j$ is lighter than $\chi_i$, as well as
{\it co-annihilation}\/ processes of the form
$\chibar_i\chi_j\to \psibar\psi$ with $i\not=j$.   
Such processes can potentially 
alter the freeze-out process in non-trivial ways.
Moreover, the existence of such processes also implies the existence of 
inelastic scattering processes of the form $\chi_i\psi \rightarrow \chi_j \psi$, 
which can potentially also impact the dynamics of freeze-out.

Second, throughout our analysis we have assumed that our dark-matter/mediator couplings $g_\chi$
are universal throughout the ensemble.  It might therefore be of interest to allow these
couplings to vary across the DDM ensemble,
thereby introducing an additional mass-dependence into the resulting cosmological abundances.
In general, the constituent-dependent dark-matter/mediator coupling $g_i$ is the coupling associated
with each constituent at its own freeze-out temperature. 
Such couplings $g_i$ will therefore experience an automatic effective ``running'' as we successively 
integrate out lighter and lighter states from the ensemble. Indeed, such running 
has already been calculated for the case in which the ensemble is a tower of KK modes~\cite{sky}.

Third, in this paper, we have restricted our analysis
to situations in which all of the DDM constituents freeze
out while non-relativistic. It is nevertheless also possible
that very light, very feebly coupled ensemble constituents
could potentially freeze out while still relativistic. This
could then significantly modify the resulting scaling 
behaviors in phenomenologically important ways.

Fourth, in this paper we have only examined the most immediate, ``zeroth-order'' constraints that
might affect our overall scaling relations.   There are, of course, many other more detailed constraints
that must be imposed before building a viable DDM model.    
These are ultimately constraints coming from potential signatures of thermal DDM scenarios within
direct-detection, indirect-detection, and collider experiments.
The case of indirect detection is particularly interesting, as 
there is the possibility of a correlation --- and even a complementarity --- between the fluxes 
of end-products from dark-matter annihilation and dark-matter decay.
While viable DDM models have been constructed~\cite{DDM2,DDMAxion} which satisfy all known experimental and observational constraints on the dark-matter sector, 
these models relied on non-thermal abundance-production mechanisms.
It still remains to determine the detailed phenomenological constraints that must be imposed
within a thermal context, and then to build actual models of this type.
Indeed, this paper represents only the first step in this direction.

Finally, it would be interesting to examine  in more detail the model-building possibilities that 
emerge from having abundances which fall as a function of mass within one part of a DDM ensemble
and rise within another, thereby experiencing different values of the scaling exponent $\gamma$ within different
regions of the ensemble.
Given that the different portions of the DDM ensemble 
can be relevant at different cosmological timescales due the variations in their masses and lifetimes,
this flexibility may enable a single DDM ensemble to simultaneously address many thorny phenomenological challenges 
that at first glance might otherwise appear to be disconnected or perhaps even contradictory.

We conclude this paper with an important comment.
Our aim throughout this paper has been to examine the scaling relations which govern a 
thermal DDM ensemble with as much generality as possible, without reference to
specific mass or energy scales.  From a phenomenological perspective, however,
given the numerous observational constraints and consistency conditions that apply to thermal freeze-out
scenarios,          
it is important to assess the natural values for the 
masses and couplings which characterize such an ensemble. 

In order to obtain a sense of the physical scales involved, we begin by recalling 
that both in canonical WIMP scenarios and in the $m_i \gg m_\phi, m_\psi$ regime 
of our DDM analysis, $\Omega_i$ is essentially determined by the ratio
$g_\chi^2 g_\psi^2/m_i^2$.  The WIMP miracle is essentially the  
observation that an abundance $\Omega_i \approx \Omega_{\mathrm{CDM}}$ is obtained 
for $m_i \approx 250$~GeV and $g_\psi = g_\chi \approx 0.65$, which yields
\begin{equation}
  \frac{m_i^2}{g_\chi^2 g_\psi^2} ~\sim~ 0.35~\tev^2~.
\label{eq:WIMPg4Overm2}
\end{equation}  
By contrast, in the $m_\phi \gg m_i, m_\psi$ regime of our DDM model, the 
corresponding quantity which determines the cross-section is 
$16 g_\chi^2 g_\psi^2 m_i^2/m_\phi^4$.  For an ensemble of particles in this 
regime, Eq.~(\ref{eq:OmegaDMLeadingOrder})
implies that 
$\Omegatot \propto \sum_i \langle \sigma_i v\rangle^{-1}$.
It therefore follows that in order to reproduce the observed value of 
$\Omega_{\mathrm{CDM}}$, such an ensemble must satisfy
\begin{equation}
  \sum_i \frac{m_\phi^4}{16 g_\chi^2 g_\psi^2 m_i^2} ~\lesssim~ 0.35~\tev^2~. 
\end{equation} 
However, by the same token, a total abundance $\Omegatot > \Omega_{\mathrm{CDM}}$ 
is problematic, signifying overproduction of dark matter.
This consideration then implies the constraint 
\begin{equation}
  \sum_i \left(\frac{m_\phi}{m_i}\right)^2 ~\lesssim~ g_\chi^2 g_\psi^2 
    \left(\frac{2.37~\tev}{m_\phi}\right)^2~.
  \label{eq:EnsembleScaleConstraint} 
\end{equation}
The masses $m_i$ and the parameters $g_\chi$, $g_\psi$, and $m_\phi$
are constrained by other considerations as well.  On the one hand, 
we have assumed that $m_\phi \gg m_i$ for all $i$.
On the other 
hand, perturbativity considerations require that $g^2_\chi, g^2_\psi \lesssim 4\pi$.  

Imposing all of these constraints, we find that the preferred regime
for a thermal DM ensemble is one in which $g_\chi$ and $g_\psi$ are large
and in which $m_i \ll m_\phi \ll \mathcal{O}(\tev)$.  However, this is easy to arrange,
for example, in scenarios in which the dark and visible sectors are 
approximately decoupled and the ensemble constituents annihilate primarily 
into other, lighter dark-sector states.  Indeed, such scenarios can accommodate
cold relic particle masses as low as $\mathcal{O}(\kev)$~\cite{LowerBoundFreezeoutMass}.
It is worth noting that
hidden-sector dark-matter models of this sort have a rich phenomenology despite
their suppressed couplings between the dark and visible sectors
(for a review, see, \eg, Ref.~\cite{FengWIMPLESREview}).    
We also note that we can always raise the mass scale of the
mediator and the dark matter simply by increasing the annihilation cross-section. 
This can be done, for example,
by adding more final states $\psi$ into the annihilation process, 
as might be arranged if $\psi$ were to carry something analogous to a color quantum number.
We also note that in the context of modified cosmologies --- for example, in
scenarios in which $\Omegatot$ is diluted by entropy injection after the freeze-out 
of the lightest constituent --- the bound in Eq.~(\ref{eq:EnsembleScaleConstraint}) 
can be considerably weakened.  Thus, in such scenarios, an even broader range of mass scales 
for thermal DDM ensembles becomes accessible.

%%%%%%%%%%%%%%%%%%%%%%%%%%%%%%%%%%%%%%%%%%%%%%%%%%%%%%%%%%%%%%%%%%%%%%%%%%%%%%%%%%%%%%

\begin{acknowledgments}

%%%%%%%%%%%%%%%%%%%%%%%%%%%%%%%%%%%%%%%%%%%%%%%%%%%%%%%%%%%%%%%%%%%%%%%%%%%%%%%%%%%%%%

We are happy to thank Kimberly Boddy and Jeff Kost for
     useful discussions.  We are also particularly grateful for
     extensive conversations with David Curtin and Adrienne Erickcek
     which helped to influence various aspects of this paper.
KRD, JK, and BT would 
like to thank the Center for Theoretical Underground 
Physics and Related Areas (CETUP$^\ast$) 
for its hospitality and partial support
during the 2016 Summer Program;  likewise, KRD and BT would also like to acknowledge
the hospitality of the Aspen Center for Physics, which is supported in part through
NSF grant PHY-1607611.
The research activities of KRD are supported in part by the Department of Energy under Grant 
DE-FG02-13ER41976 (DE-SC0009913)
and by the National Science Foundation through its employee IR/D 
program.  The research activities of JK are supported in part by NSF CAREER grant PHY-1250573, and  
those of BT are supported in part by NSF grant PHY-1720430.
The opinions 
and conclusions expressed herein are those of the authors, and do not represent 
any funding agencies. 
 
\end{acknowledgments}  

%%%%%%%%%%%%%%%%%%%%%%%%%%%%%%%%%%%%%%%%%%%%%%%%%%%%%%%%%%%%%%%%%%%%%%%%%%%%%%%%%%%%%%

\end{document}